\documentclass[fleqn,usenatbib]{mnras}

\usepackage{newtxtext,newtxmath}

\usepackage[T1]{fontenc}

\DeclareRobustCommand{\VAN}[3]{#2}
\let\VANthebibliography\thebibliography
\def\thebibliography{\DeclareRobustCommand{\VAN}[3]{##3}\VANthebibliography}

\usepackage{graphicx}	
\usepackage{amsmath}	
\usepackage{amssymb}	
\usepackage{booktabs}   
\usepackage{multirow}   
\usepackage{comment}    
\usepackage{gensymb}   

\title[Galaxy fundamental scaling relations]{The Molecular-Gas Main Sequence and Schmidt-Kennicutt relation are fundamental, the Star-Forming Main Sequence is a (useful) byproduct}

\author[W. M. Baker et al.]{
William M. Baker$^{1,2}$\thanks{E-mail: wb308@cam.ac.uk},
Roberto Maiolino$^{1,2,3}$,
Francesco Belfiore$^{4}$,
Asa F. L. Bluck$^{5}$,
Mirko Curti$^{1,2}$,
\newauthor{}
Dominika Wylezalek$^{6}$,
Caroline Bertemes$^{6}$,
M. S. Bothwell$^{7}$,
Lihwai Lin$^{8}$,
Mallory Thorp$^{9}$,
Hsi-An Pan$^{10}$
\\
$^{1}$Kavli Institute for Cosmology, University of Cambridge, Madingley Road, Cambridge, CB3 0HA, UK\\
$^{2}$Cavendish Laboratory - Astrophysics Group, University of Cambridge, 19 JJ Thomson Avenue, Cambridge, CB3 0HE, UK\\
$^{3}$Department of Physics and Astronomy, University College London, Gower Street, London WC1E 6BT, UK\\
$^{4}$INAF— Osservatorio Astrofisico di Arcetri, Largo E. Fermi 5, I-50125, Florence, Italy\\
$^{5}$Department of Physics, Florida International University, 11200 SW 8th Street, Miami, FL, USA\\
$^{6}$Zentrum für Astronomie der Universität Heidelberg, Astronomisches Rechen-Institut, Mönchhofstr, 12-14 69120 Heidelberg, Germany\\
$^{7}$ Institute of Astronomy, University of Cambridge, Madingley Road, Cambridge CB3 0HA, UK\\
$^{8}$Institute of Astronomy \& Astrophysics, Academia Sinica, Taipei 10617, Taiwan\\
$^{9}$Department of Physics \& Astronomy, University of Victoria, Finnerty Road, Victoria, British Columbia, V8P 1A1, Canada\\
$^{10}$ Department of Physics, Tamkang University, No.151, Yingzhuan Road, Tamsui District, New Taipei City 251301, Taiwan 
}

\date{Accepted XXX. Received YYY; in original form ZZZ}

\pubyear{2015}

\begin{document}
\label{firstpage}
\pagerange{\pageref{firstpage}--\pageref{lastpage}}
\maketitle

\begin{abstract}
We investigate the relationship between the star formation rate (SFR), stellar mass ($M_*$) and molecular gas mass ($M_{H_2}$) for local star-forming galaxies. We further investigate these relationships for high-z (z=1-3) galaxies and for the hosts of a local sample of Active Galactic Nuclei (AGN). We explore which of these dependencies are intrinsic and which are an indirect by-product by employing partial correlation coefficients and random forest regression. We find that for local star-forming galaxies, high-z galaxies, and AGN host galaxies, the Schmidt-Kennicutt relation (SK, between $M_{H_2}$ and SFR), and the Molecular Gas Main Sequence (MGMS, between $M_{H_2}$ and $M_*$) are intrinsic primary relations, while
 the relationship between $M_*$ and $SFR$, i.e. the Star-Forming Main Sequence (SFMS), is an indirect by-product of the former two.
Hence the Star-Forming Main Sequence is not a fundamental scaling relation for local or high-redshift galaxies. We find evidence for both the evolution of the MGMS and SK relation over cosmic time, where, at a given stellar mass, the higher the redshift, the greater the molecular gas mass and the star formation efficiency. 
We offer a parameterisation of both the MGMS and SK relation's evolution with redshift, showing how they combine to form the observed evolution of the SFMS. 
 In addition, we find that the local AGN host galaxies follow an AGN-MGMS relation (as well as a AGN-SK relation), where the MGMS is offset to lower $M_{H_2}$ for a given $M_*$ compared to local SF galaxies.

\end{abstract}




\section{Introduction}

The fundamental properties of galaxies tracing their cycle of star formation, star formation rate (SFR), molecular gas mass ($M_{H_2}$) and stellar mass ($M_*$), are connected by three scaling relations.
The Schmidt-Kennicutt (SK) relation \citep{Schmidt1959, Kennicutt1998}, illustrates how, as the amount of molecular gas in a galaxy increases, the star formation rate also increases, and is usually considered (to a first order), simply molecular gas being the fuel for star formation. The Molecular Gas Main Sequence \citep[MGMS][]{Lin2019}, describes how an increase in molecular gas mass is correlated to an increase in stellar mass, likely resulting from M$_*$ tracing the gravitational potential, retaining gas and/or as a consequence of higher pressure, fostering the conversion of HI to H$_2$. Finally, the Star Forming Main Sequence \citep[SFMS][]{2004MNRASBrinchmann, Renzini+PengMS2015ApJ...801L..29R,Whitaker2012ApJ...754L..29W, Sandles2022arXiv220706322S} describes the observed correlation between the star formation rate and the stellar mass.

These scaling relations have been explored on both global and resolved scales 
\citep[examples of studies include][]{2004MNRASBrinchmann, NOESKE_SFMS_2007ApJ...660L..43N,Whitaker2012ApJ...754L..29W, 2017BolattoEDGEApJ...846..159B, 2017UtomoApJ...849...26U, Lin2019, Lin2020, Ellison2021AlmaQuest5, Pessa2021A&A...650A.134P, Sanchez2021MNRAS.503.1615S, Baker2022MNRAS.510.3622B}
In particular, using spatially resolved data from ALMaQUEST and MaNGA, \citet{Lin2019} found that, on resolved scales, the resolved Schmidt-Kennicutt (rSK) relation is the most strongly correlated and with the lowest dispersion, followed closely by the resolved molecular gas main sequence (rMGMS), whilst the resolved star forming main sequence (rSFMS) appeared to have the largest scatter of the three. They suggested that the rSFMS could be a by-product of the SK and MGMS relations. This would be the case when both SFR and stellar mass are directly correlated to a third variable, i.e. the molecular gas mass, and hence are indirectly correlated themselves. These findings were further investigated in \cite{Ellison2021AlmaQuest5}, who traced these relations for individual galaxies.

These results were explored more in detail in \citet{Baker2022MNRAS.510.3622B}, where we used partial correlation coefficients and random forest regression to investigate the intrinsic and induced dependencies between $\Sigma_{SFR}$, $\Sigma_{H_2}$ and $\Sigma_*$, as well as their scaling relations. We found that the rSK relation and rMGMS relations are intrinsic, direct and fundamental relations, while the rSFMS is simply a by-product of the two former relations. In other words, the resolved star formation rate has no fundamental dependence on stellar mass surface density and the rSFMS is simply a consequence of both $\Sigma_{SFR}$ and $\Sigma_*$ being directly correlated with $\Sigma_{H_2}$.

This naturally leads to speculation about whether the corresponding integrated (i.e. galaxy wide) scaling relations have the same inter-dependencies. Is the global SFMS simply a by-product of the global SK relation and MGMS? 
In addition, is this still the case at high-z or do the scaling relations themselves change their dependencies over cosmic time?

Research into integrated scaling relations at both low and high-z, that involve molecular gas mass, have generally focused on derived quantities such as depletion time (the inverse of the star formation efficiency, SFE=SFR/M$_{H_2}$) or gas fraction (f$_{gas}$=M$_{H_2}$/M$_*$) \citep[][]{Saintonge2016MNRAS.462.1749S,Tacconi_PHIBBS_2018ApJ...853..179T, Tacconi2020, Saintonge2022arXiv220200690S, Dou_Peng_2021ApJ...907..114D}.
While convenient for many aspects, the use of these combined and derived quantities (i.e. not directly observed) makes it tricky to disentangle their intrinsic dependencies as they are themselves a combination of other directly observed quantities (SFR, $M_*$ and $M_{H_2}$). Furthermore, many of the past studies explore these quantities as a function of the location of galaxies relative to the star forming main sequence, which in itself may be indirectly caused by the star formation rate and stellar masses' individual correlations with molecular gas mass, as it is for the resolved quantities \citep[][]{Lin2019,Baker2022MNRAS.510.3622B}.

A further interesting question is what scaling relations look like for galaxies containing Active Galactic Nuclei (AGN). In the past, studies have shown that AGN appear to populate the Green Valley region, i.e. hosted in galaxies with reduced star formation rates relative to galaxies on the SFMS 
\citep[][]{2008SilvermanApJ...675.1025S}. A common explanation of this is that AGN feedback causes a decrease in the SFR via expulsion of gas or that it warms the cold gas reducing star formation efficiency \citep{Cimatti, 2017MahoroMNRAS.471.3226M}. 
Other studies have found that, when accounting for stellar mass and redshift, galaxies with AGN have higher star formation rates (by the order of 3 to 10) compared to those galaxies comparable in stellar mass and redshift without an AGN \citep[][]{Shimizu2017MNRAS.466.3161S, Florez2020MNRAS.497.3273F}. So there are apparently contrasting results on this topic.
However, many of these results might be inconclusive because, again, they deal with inter-correlated quantities and may fail to disentangle primary correlations from secondary, induced dependences. As evidence of this potential problem, by using contemporary hydrodynamical cosmological simulations, \cite{Piotrowska2022MNRAS.512.1052P} found no significant correlation between the black hole accretion rate $\dot M_{BH}$, i.e. the bolometric AGN luminosity, and galaxy quenching. 
 Similarly, \cite{Ward_AGNFeedback_2022MNRAS.514.2936W} showed that there is not necessarily an anti-correlation between the instantaneous AGN activity and star-formation rate even in simulations where the AGN is demonstrably causally responsible for quenching.

In this paper we tackle these issues by building upon the work of \citet{Baker2022MNRAS.510.3622B} for local star-forming galaxies, using partial correlation coefficients and random forest regression to investigate the scaling relations on integrated (i.e. galaxy wide) scales. We do this by utilising a variety of surveys that obtained measurements of the CO(1-0) or CO(2-1) lines, combined with measurements of SFR and $M_*$ from the Sloan Digital Sky Survey SDSS DR7 MPA-JHU catalogue \citep[][]{2004Tremonti, 2004MNRASBrinchmann}. This enables us to probe whether the findings on resolved scales are mirrored on integrated scales.

In Section \ref{sec:data} we introduce the data, explain our selection criteria and how we derive the required physical quantities.

Section \ref{sec:Statistical methods} describes the analysis tools we use to probe the intrinsic dependencies of the scaling relations. We introduce both partial correlation coefficients and random forest regression and explain how they enable us to disentangle the dependencies between multiple inter-correlated quantities

Section \ref{sec:localSF} explores the scaling relations for local star-forming galaxies by using partial correlation coefficients and random forest regression. We also parameterise a MGMS for our star-forming population of galaxies in the local universe (and what effects the inclusion/exclusion of upper limits has on the results). 

In Section \ref{sec:high-z}, using the same methods, we utilise the PHIBBS \citep[][]{Tacconi_PHIBBS_2018ApJ...853..179T} sample to explore redshift evolution of the scaling relations for galaxies in the redshift range 0.5<z<4.5. We also split the sample into redshift bins in order to investigate possible evolution of the MGMS, SK and SFMS. We then offer a parameterisation of the observed evolution. 

In Section \ref{sec:AGN} we investigate the scaling relations for local AGN using the BAT-CO \citep[][]{2021KossBATApJS..252...29K} survey, by applying the same techniques to find out if AGN follow similar scaling relations. We are also able to explore whether the AGN luminosity contributes to determining the star formation rate of the host galaxy. This gives us an indication of the relative importance of instantaneous AGN feedback (as probed by AGN X-ray luminosity) in affecting star formation.

\section{Data, selection criteria, and physical quantities}

\label{sec:data}

For the local star-forming galaxies sample, we use data from a combination of surveys that obtained integrated measurements of either the CO(1-0) line or the CO(2-1) line.

\begin{table*}
\caption{Table listing the different samples used, how many galaxies they contain, the number of galaxies that make it past our cuts, and the primary method used to obtain the molecular gas masses. The first part of the table contains the surveys we combined to make our local star-forming galaxy sample. It can be seen the majority of our local star forming galaxies are drawn from either MASCOT or xCOLD GAS.
The second part of the table contains the local AGN sample drawn from BAT. The third contains the PHIBBS sample from which we select our high-z galaxies. The PHIBBS sample contains molecular gas mass measurements from CO and dust masses.}
\centering
\label{table:surveys}
\begin{tabular}{l c c c c}
\toprule
\multirow{2}{*}{Survey}  &   \multirow{2}{*}{\# of galaxies} &   \multirow{2}{*}{\# of galaxies used} & \multirow{2}{*}{ (primary) transistion}  \\[+5pt]

\midrule

xCOLD GASS & 532 & 217 & CO(1-0)\\
      
MASCOT & 187 & 119 & CO(1-0)   \\

ALLSMOG & 97 & 35 & CO(2-1)  \\

EDGE & 126 &  17 & CO(1-0) \\

HERSCHEL-SDSS Stripe82 & 78 & 35  & CO(1-0) \\

ALMaQUEST & 46  & 20  & CO(1-0) \\

\midrule
BAT & 213  & 208  & CO(2-1) \\
\midrule
PHIBBS (sample) &  1309 & 824  & Combination \\

\bottomrule
\end{tabular}
\end{table*}

Table \ref{table:surveys} lists the different surveys used and includes information on how many galaxies they include, how many make it past our selection criteria (discussed in Section \ref{sec:selection}), and the primary CO transition used for determining the molecular gas masses. In addition, in this section we also include brief subsections giving a quick overview of each survey. We will then provide an explanation of how galaxies were selected and how the physical quantities were derived in a homogeneous way.

\subsection{Databases}

\subsubsection{MASCOT}
The MaNGA-ARO Survey of CO Targets (MASCOT) \citep{DominikaMascott2022MNRAS.510.3119W} is a single-dish European Southern Observatory survey which is being conducted at the Arizona Radio Observatory. To date the survey has observed 187 local star-forming galaxies selected from MaNGA (Mapping Nearby Galaxies at Apache Point Observatory) \citep[][]{Bundy2015} and obtained integrated measurements of the CO(1-0) emission line.

\subsubsection{ALLSMOG}
The APEX low-redshift legacy survey for molecular gas (ALLSMOG) \citep{2017CiconeALLSMOGA&A...604A..53C} is a survey conducted with the APEX telescope, which observed 88 local star-forming galaxies, which were selected from SDSS, obtaining integrated CO(2-1) measurements. They also obtained IRAM measurements of 9 additional galaxies in both CO(1-0) and CO(2-1) increasing the total survey size to 97 galaxies.

\subsubsection{xCOLD GASS}
The xCOLD GASS survey \citep{2017XCOLDGASSAintongeApJS..233...22S} builds upon the COLD GASS survey \citep[][]{Saintonge_COLDGAS_2011MNRAS.415...32S} and contains CO(1-0) measurements for 532 local galaxies selected from SDSS. The measurements were taken with the IRAM 30m telescope and targeted galaxies with stellar masses larger than $10^9 M_{\odot}$.

\subsubsection{EDGE}
The Extragalactic Database for Galaxy Evolution survey (EDGE) \citep{2017BolattoEDGEApJ...846..159B} conducted observations of CO(1-0) for 126 local galaxies selected from the Calar Alto Legacy Integral Field Area (CALIFA) \citep[][]{Sanchez_CALIFA_2012A&A...538A...8S} using the Combined Array for Millimeter-wave Astronomy (CARMA).

\subsubsection{HERSCHEL-SDSS Stripe 82}
We use CO(1-0) data from 78 star forming galaxies that were part of a survey used in \citet{2018BertemesMNRAS.478.1442B}. The CO data were obtained using the IRAM 30m telescope.

\subsubsection{ALMaQUEST}
The ALMaQUEST survey \citep{Lin2020} provides CO(1-0) ALMA maps for 46 local galaxies selected from MaNGA. The galaxies span a range of types including, starburst, main sequence, and green valley.  For further information see \cite{Lin2020}.

\subsubsection{BAT}
The BAT (Burst Alert Telescope) CO survey \citep[][]{2021KossBATApJS..252...29K} obtained CO(2-1) (230.538GHz) emission line measurements for 213 local AGN galaxies, which were selected from the Swift-BAT AGN Spectroscopic Survey \citep[BASS][]{2017KossBASSApJ...850...74K}. The AGN were selected via hard X-ray emission (14-195keV), and with redshifts in the range of 0.001<z<0.05 (therefore the sample is defined without selection on star formation activity so can include SF, green valley and quiescent galaxies.). We use the molecular gas mass ($M_{H_2}$), combined with stellar masses ($M_*$) and SFRs (obtained via IR luminosity measurements) from \citet{2021KossBATApJS..252...29K}, combined with the absorption corrected hard X-ray luminosity in the 2-10keV range from \citet{2017KossBASSApJ...850...74K}. 

\subsubsection{PHIBBS}
In order to explore high-redshift galaxies in this analysis we incorporate the PHIBBS collection of galaxy samples \citep{Tacconi_PHIBBS_2018ApJ...853..179T}.  The \citet{Tacconi_PHIBBS_2018ApJ...853..179T} sample contains 1309 measurements of $M_{H_2}$ for galaxies with redshifts in the range 0<z<4, this means the galaxies in the sample span the majority of the age of the Universe. These galaxies range in stellar masses from $10^9M_\odot<M_*<10^{10.9}M_\odot$ with a wide range of star formation rates. The molecular gas masses are obtained via a combination of different methods. There are 667 CO line measurements which are converted first into CO luminosities following the methodology of \citet{Solomon1997}, then into molecular gas masses using a metallicity-dependent conversion factor as explained in \cite{Genzel2015ApJ...800...20G}. 512 molecular gas measurements are obtained from dust masses which are inferred from Herschel far-infrared SEDs and 130 were obtained single photometric point measurement at $\sim$1mm, assuming a fixed dust temperature and the method described in \cite{Scoville2017ApJ...837..150S}. A metallicity-dependent dust to gas conversion factor was used. For more details see \cite{Genzel2015ApJ...800...20G,Tacconi_PHIBBS_2018ApJ...853..179T}.

\subsection{Selection of galaxies}
\label{sec:selection}

We focus on star forming galaxies selected via their location on the
[NII]-BPT \citep[][]{Baldwin} diagram based on the \citet{2003Kauffmann} dividing line.
This selection is needed for estimating the metallicity when using the metallicity dependent conversion factor (as strong-line metallicity diagnostics are only calibrated for star forming galaxies) and also to ensure that the determination of the SFR is not affected by the presence of an AGN or LI(N)ER-like (low ionisation (nuclear) emitting region) emission. 
Nebular line fluxes for such selection were obtained from the SDSS DR7 MPA-JHU catalogue\footnote{https://wwwmpa.mpa-garching.mpg.de/SDSS/DR7/}  We drop galaxies not included in the catalogue. The exploration of the trends for AGN is obviously interesting, but we explore this via the more carefully selected and more complete, BAT-AGN sample, as discussed later on. We also note that the BPT-SF selection removes most of the quiescent galaxies, but it does not introduce a sharp cut for SF galaxies, allowing for a larger dynamic range in the exploration of SF galaxies than what a sSFR or $\Delta$MS cut would do.

In addition, except when considering the maximum likelihood analysis, we exclude all CO non-detections and upper limits from our analysis.
This of course reduces our sample size, but still gives us 443 galaxies in the local sample, enough to enable us to use partial correlation coefficients and random forest regression (provided we limit the number of features evaluated). However, we also consider the case of including upper limits to explore the robustness of our results. This will be discussed more in detail in the next sections.

For the local AGN analysis, we use all the galaxies included in the BAT AGN sample \citep{2021KossBATApJS..252...29K} which also have X-ray luminosities from the BASS sample \citep{2017KossBASSApJ...850...74K}. This gives us a total of 208 AGN host galaxies.

In the high-z analysis, we use all the PHIBBS galaxies that have a redshift greater than 0.5, giving us 824 galaxies. We note that, when binning by redshift for parameterising the possible evolution of the MGMS,  we also include the local PHIBBS galaxies (z<0.5) to provide a consistent local reference sample, this gives us a combined total of 1309 galaxies. 

\subsection{Molecular gas masses, stellar masses and star formation rates}
\label{sec:methods}
In order to ensure consistency between samples we convert CO(2-1) luminosity into CO(1-0) luminosity, assuming $\rm L_{C0(1-0)}=L_{C0(2-1)}/0.8$ \citep[][]{2017CiconeALLSMOGA&A...604A..53C}.

We convert from CO flux into CO luminosity using the well known relation from \citet{Solomon1997}
\begin{equation}
   \rm L'_{CO}= 3.25\times10^7\,S_{CO}\,\Delta V\, \nu_{obs}^{-2}\, D_L^2\, (1+z)^{-3}\, 
\end{equation}
where $\rm L'_{CO}[K\,km\,s^{-1}\,pc^2]$ is the CO line luminosity, $\rm S_{CO}\Delta V[Jy\,km\,s^{-1}]$ is the velocity integrated flux, $\rm \nu_{obs}[GHz]$ is the observed line frequency, and $\rm D_L[Mpc]$ is the distance luminosity.

We then convert from CO luminosity into molecular gas mass by using the conversion factor $\rm \alpha_{CO}=4.35 M_\odot\,(\rm K km/s pc^2)^{-1}$ \citep[][]{Bolatto2013}. We also use a metallicity-dependent conversion factor of the form $\rm \alpha_{ CO}=4.35\bigg(\frac{Z}{Z_\odot}\bigg)^{-1.6} M_\odot (K\, km\,/s\, pc^2)^{-1}$ \citep[][]{Accurso2017MNRAS.470.4750A} in order to test the difference and find no significant change to our results. This is unsurprising as this mirrors what already was found in \citet{Lin2019} and \cite{Baker2022MNRAS.510.3622B}. Nevertheless partial correlation coefficients themselves can be influenced strongly by the conversion factor in certain scenarios, in particular involving metallicity \citep[][]{Baker_rFMR_2022arXiv221003755B}.

For the local star-forming galaxies, obtained from a combination of surveys, we use stellar masses and star-formation rates from the MPA-JHU catalogue. We convert all stellar masses and star formation rates to a \cite{Chabrier_IMF_2003PASP..115..763C} Initial Mass Function (IMF) to ensure consistency. 
As part of our analysis we also include the bulge to total ratio (B/T) as measured by \cite{Mendel2014ApJS..210....3M}, who fit a bulge and disk component to each galaxy then take B/T to be the mass of the bulge divided by the total mass, i.e. that of the bulge plus the disk.

For the local AGN we use the molecular gas masses, stellar masses, star formation rates and AGN luminosities provided in \citet{2021KossBATApJS..252...29K, 2017KossBASSApJ...850...74K}.
For the high-z analysis we use the molecular gas masses, stellar masses and  star formation rates from PHIBBS \citep[][]{Tacconi_PHIBBS_2018ApJ...853..179T}.
We note that for the AGN and the PHIBBS surveys, as we use their provided molecular gas masses, stellar masses and star formation rates, these have been calculated in different ways and under varying assumptions. However, due to the fact we are considering each of these samples independently we do not expect this to cause a significant issue: what is important is that consistency is maintained within each sample.

\section{Statistical methods}

\label{sec:Statistical methods}

\subsection{Partial Correlation coefficients}
How can one find the intrinsic correlation between two quantities whilst taking into account possible dependencies on further quantities? Partial correlation coefficients enable us to take the correlation between two quantities whilst holding fixed a third (or more) constant, thereby controlling for these additional quantities. This provides a valuable test of whether a correlation is intrinsic or rather a by-product of other more fundamental dependencies.
The partial correlation coefficient between two quantities 1 and 2 whilst controlling for a third quantity, 3, is given by
\begin{equation}
    \rho_{12|3}=\frac{\rho_{12}-\rho_{13}\rho_{23}}{\sqrt{1-\rho_{13}^2}\sqrt{1-\rho_{23}^2}}.
\end{equation}

The values of the partial correlation coefficients can be used to define a gradient arrow \citep{2020Bluck} in a 3D diagram where quantities 1 \& 2 are on the x and y axes and quantity 3 is colour-coded (i.e. on the z axis).
The gradient arrow points in the direction of greatest increase in the third colour-coded quantity. This enables the strength of partial correlation coefficients between the colour-coded quantity, and each of those on the $x$ and $y$ axis, to be shown graphically. 
The arrow angle is given by
\citep[adapting equation from][]{2020Bluck}
\begin{equation}
    \text{tan}(\theta)=\frac{\rho_{13|2}}{\rho_{23|1}}
    \label{theta}
\end{equation}
where $\theta$, the angle, is measured from the positive $y$ axis, like a bearing, in a clockwise direction. Another key strength of partial correlation coefficients is that they explain the direction of the correlation, i.e. whether quantities are correlated or anti-correlated. However, partial correlation coefficients do require monotonic trends, this is why we also use random forest regression which can also probe non-monotonic relationships.

\subsection{Random Forest Regression}

A random forest is an amalgamation of many different decision trees and is a form of supervised (identifiable labels) machine learning. The criteria for each split in the decision tree is to minimise Gini Impurity, which is a measure of how accurate the split is in the decision trees. From this, the random forest can identify which of the parameters are the most important in determining a target quantity. With a normalised performance score for each parameter, one can directly compare several different parameters (which can be investigated simultaneously) and determine which achieves the best performance in determining the target. 

The random forest is trained on 80\% of the available data and then applied to the remaining 20\% to obtain the importances. The mean squared error obtained by fitting the training and test samples is compared to ensure that the random forest is not overfitting, i.e fitting the noise of the training sample which would negatively impact its performance when applied to new data sets. We used a full grid search to tune the hyper-parameters of our random forest, which consist of parameters such as the minimum number of samples in an end node, or the number of trees in the forest. 
A primary advantage of random forest regression is that it can find non-monotonic trends in the data - something that partial correlation coefficients, with their requirement of monotonicity, cannot do. It also explores all the parameters simultaneously and has been found to be able to uncover the intrinsic dependence of a given quantity \citep[for more details about random forests and their use in determining causation see][]{2020Bluck, 2020BluckB,Bluck2022A&A...659A.160B, Piotrowska2022MNRAS.512.1052P}.

\section{Results: Local Star forming galaxies}

\label{sec:localSF}

\begin{figure*}
    \centering
    \includegraphics[width=\columnwidth]{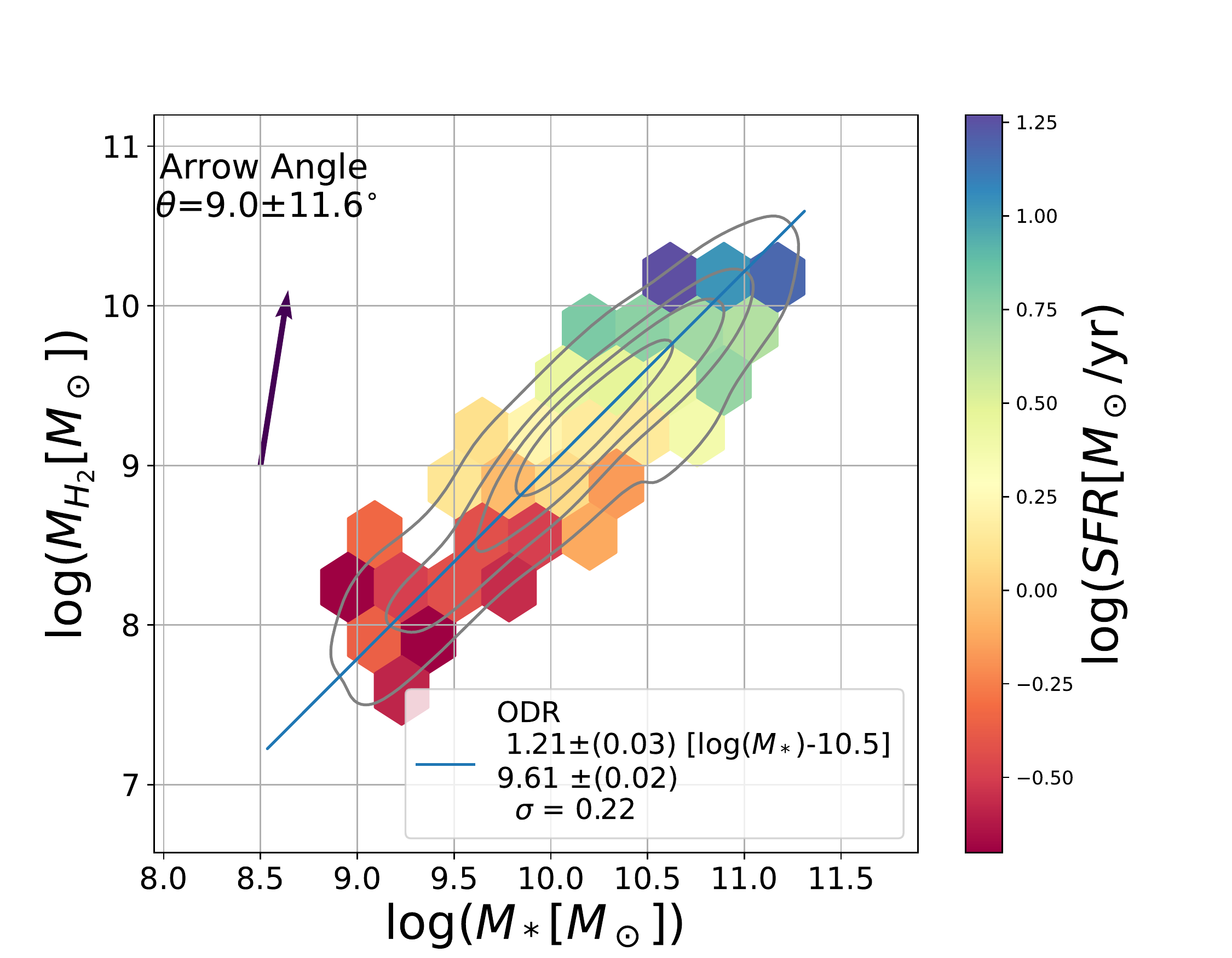}
    \includegraphics[width=\columnwidth]{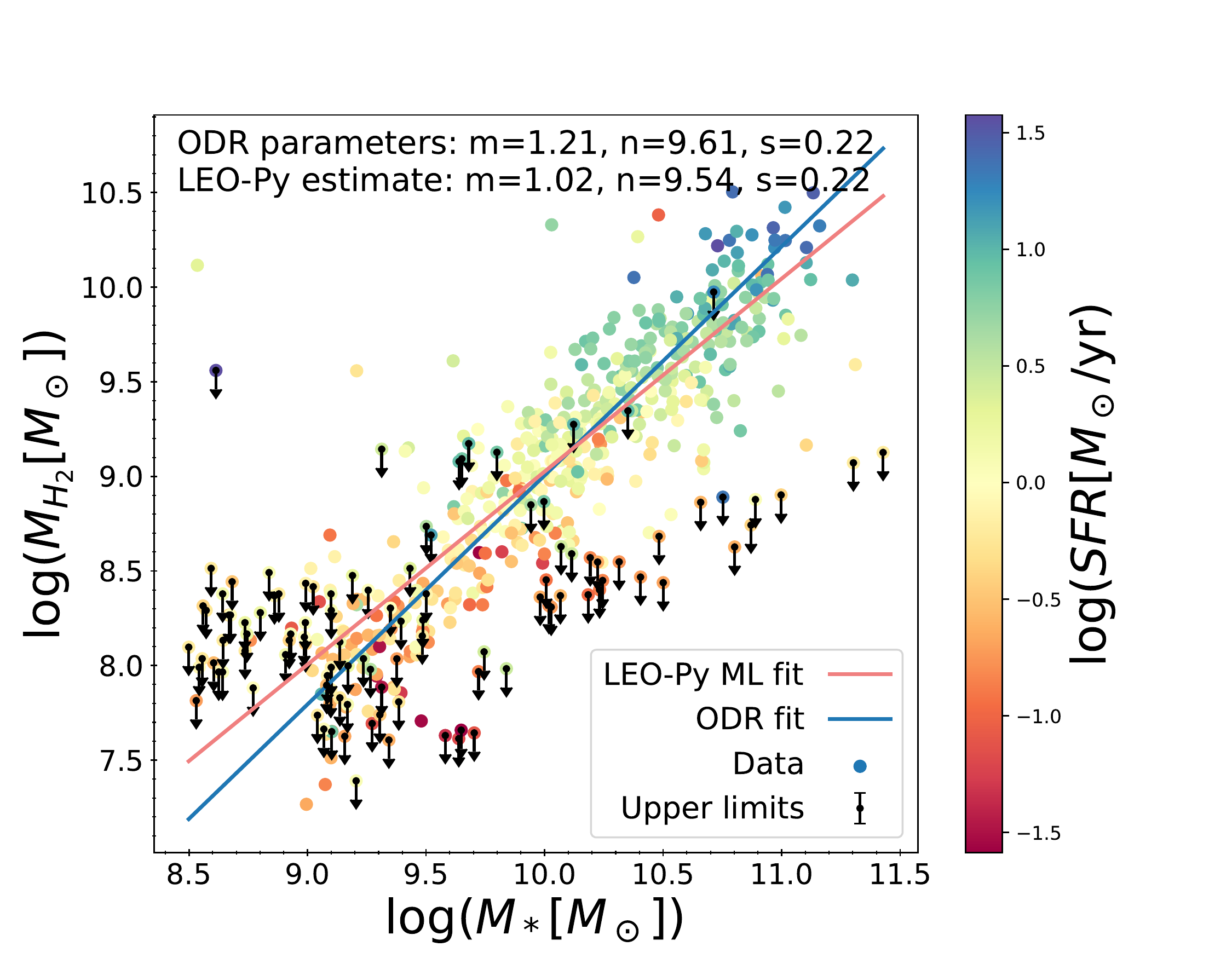}
    \caption{Molecular gas mass (M$_{H2}$) versus stellar mass (M$_{*}$), colour-coded by star formation rate (SFR), for local star forming galaxies. Left: hexagonal binned (2D histogram) in which each bin is colour-coded by the mean log star formation rate in in the bin. The contours show the distribution of galaxies, where the outer density contour is set to include 90\% of the population.  The best-fit blue line is calculated via orthogonal distance regression and reveals a tight Molecular Gas Main Sequence relation in the local universe with a scatter of $\sigma=0.22$dex. The partial correlation coefficient arrow points in the direction of the largest increasing gradient in SFR (color-coded quantity). The arrow shows that the SFR is almost completely determined by the molecular gas mass (arrow consistent with being vertical) with little to no contribution from the stellar mass (at fixed M$_{H2}$). This indicates that the SFMS is not a fundamental scaling relation. 
    The right panel shows the distribution of the individual galaxies, including upper limits on M$_{H2}$.
   The blue line is the ODR fit obtained by fitting to (only) the detections (as in the left panel) whilst the orange line is the fit obtained by Maximum Likelihood analysis, where the Likelihood takes into account upper limits. There are only slight differences between the two fits.}
    \label{fig:Mstar-H2-SFR-Starforming}
\end{figure*}

\begin{figure}
    \centering
    \includegraphics[width=\columnwidth]{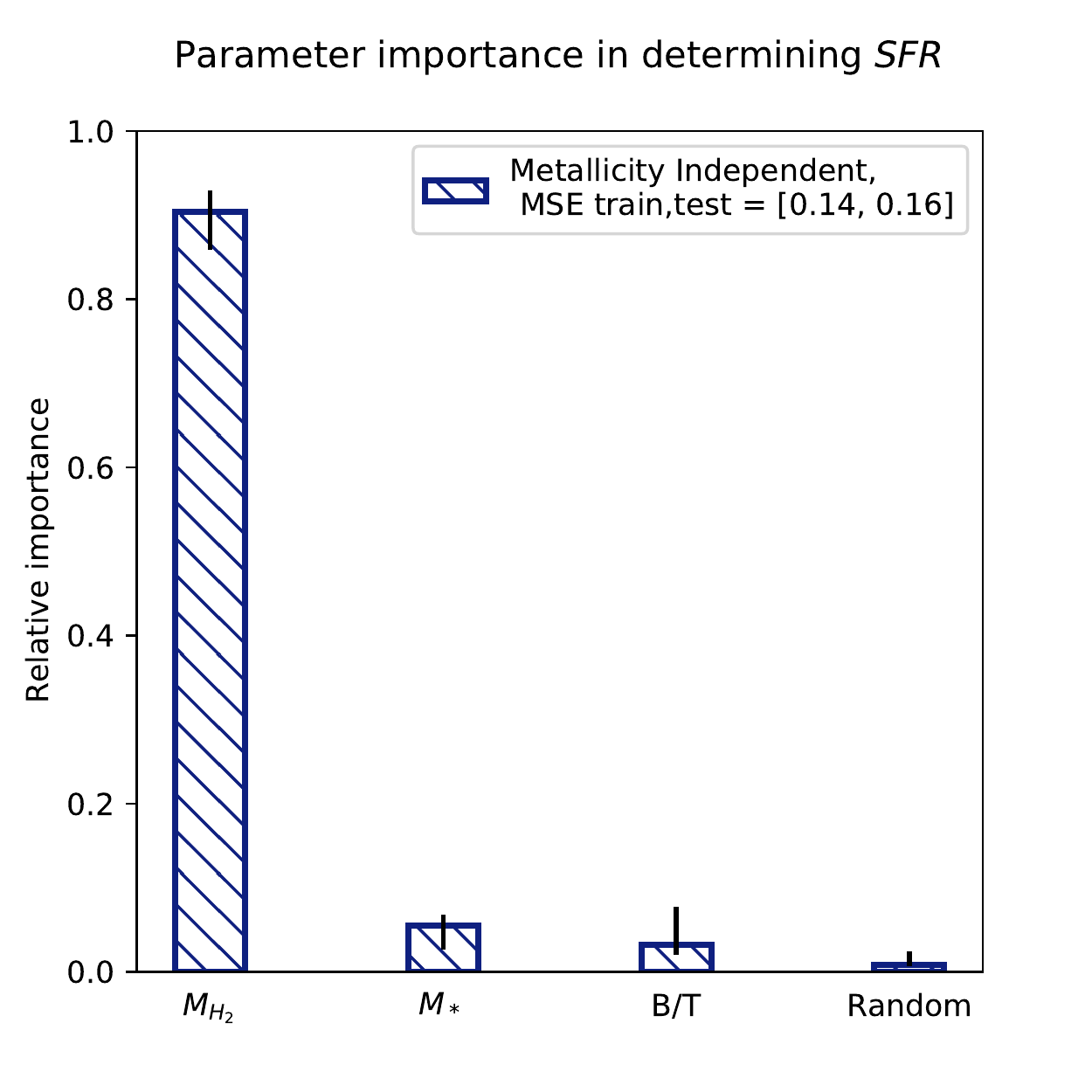}
    \caption{Random forest regression exploring the importance of galactic parameters for determining the SFR of local star forming galaxies. The parameters under consideration are: molecular gas mass $M_{H_2}$, stellar mass $M_*$, bulge-to-total ratio (B/T), and a uniform random variable (Random). The errors are obtained by bootstrap random sampling 100 times.
    SFR is almost entirely determined by the molecular gas mass whilst the other parameters are consistent with the uniform random variable. This confirms that the SFMS is insignificant compared to the Schmidt-Kennicutt relation in driving the SFR of local star-forming galaxies and that the SK relation does not depend on $M_*$ or B/T. }
    \label{fig:RF_starforming}
\end{figure}

We start by exploring the scaling relations for local star-forming galaxies. To do this we investigate which quantity out of molecular gas mass and stellar mass is more important in driving the star formation rate. This will inform us which of the Schmidt-Kennicutt and the Star Forming Main Sequence is the more fundamental relation.

The top-left panel of Figure \ref{fig:Mstar-H2-SFR-Starforming} shows a 2D histogram of the three quantities in the form of $M_{H_2}$ against $M_*$ colour coded by SFR. The colour-coding corresponds to the mean value of the log SFR of the galaxies within each bin. There are a minimum of two galaxies per bin. The density contours show the distribution of galaxies on the $M_{H_2}$-$M_*$ plane, with the outer contour set to include 90\% of the population. 
The partial correlation coefficient arrow points in the direction of the greatest increasing gradient in the colour-coded quantity (SFR).
As can be seen from the color gradient and from the gradient arrow, the molecular gas mass is almost totally responsible in determining the SFR compared to the stellar mass (the arrow angle is consistent with 0$^\circ$). This shows that the SFR is essentially totally driven by the molecular gas content, while the stellar mass does not play any role in determining the SFR once the dependence on M$_{H2}$ is taken into account. In other words the Schmidt-Kennicutt relation is far more important than the Star Forming Main Sequence in determining SFR.

The contours also illustrate that galaxies occupy a tight relation on the $M_{H_2}$-$M_*$ plane, i.e. the integrated version of the Molecular Gas Main Sequence (MGMS). The scatter of the relation is fairly small,
0.22dex, which is comparable to the scatter of the Star Forming Main Sequence ($\approx$0.3) and probably even smaller if one considers that the observational uncertainties on the molecular gas measurements are significantly larger than those of the SFR.

We fit the MGMS with the following linear relation

\begin{equation}
    \rm log(M_{H_2})=m\,[log(M_*)-10.5]+log(M_{H2})_{10.5}
     \label{eq:odr_general}
\end{equation}
where m is the gradient of the line and $\rm log(M_{H2})_{10.5}$ gives the value of $\rm log(M_{H_2})$ at $\rm log(M_*/M_\odot)=10.5$.
The blue line in top-left panel of Fig.~\ref{fig:Mstar-H2-SFR-Starforming} shows the best-fit calculated using orthogonal distance regression (ODR) to the M$_{H2}$--M$_*$ relation. This means it takes into account uncertainties in both the $x$ axis quantity ($M_*$) and the $y$ axis quantity $M_{H_2}$. 
The resulting MGMS for star-forming galaxies in the local universe inferred by us is given by $m=1.21$ and $\rm log(M_{H2})_{10.5}=9.61\, (\pm0.02)$, i.e.
\begin{equation}
    \rm log(M_{H_2})=1.21\,(\pm 0.03)\,[log(M_*)-10.5]\,+\,9.61 (\pm 0.02)
    \label{eq:odr_local}
\end{equation}

We next investigate the impact of the exclusion of upper limits. 
In the right of Figure  \ref{fig:Mstar-H2-SFR-Starforming} we show the distribution of individual galaxies, again color-coded by their SFR, and also include upper limits to explore their effect on the slope of the correlation between M$_{H2}$ and M$_*$, i.e. the MGMS.
The blue line is the same orthogonal distance regression fit given in Equation \ref{eq:odr_local}. This means it is only calculated for the detections and excludes the impact of any upper limits. It is, however, possible to include upper limits in the calculation of a line of best fit by incorporating them into the likelihood. This likelihood can then be maximised to find the best-fit parameters. To incorporate the upper limits into the likelihood we make use the LEOPY Python module \citep[][]{Leopy_Feldmann_2019A&C....2900331F}. The orange line  gives the maximum likelihood estimate of the best-fit parameters for the model. It can be seen that the ODR approach and the maximum likelihood approach are very similar.
This means that our ODR fiting approach to defining the MGMS is not biased significantly by the exclusion of upper limits. 

We then use the Random Forest regression analysis to investigate the importance of galactic parameters in predicting the SFR.
Figure \ref{fig:RF_starforming} is a bar-chart showing the results of the random forest regression parameter importance in determining the star formation rate. Specifically, it explores which of molecular gas mass ($M_{H_2}$), stellar mass ($M_*$), or bulge-to-total (B/T) is most important in driving SFR (a random variable is also included to control that it gives approximately zero importance). The errors are obtained by bootstrap random sampling 100 times.
Clearly the SFR is almost entirely determined by the molecular gas mass, whilst all other parameters considered (including M$_*$) are consistent with the uniform random variable, i.e. no contribution above random. This supports the result found in Figure \ref{fig:Mstar-H2-SFR-Starforming}, i.e. that {\it the SFR is determined by the molecular gas mass (SK relation) and has little to no intrinsic dependence on the stellar mass once the dependence on $M_{H_2}$ is taken into account, hence the star forming main sequence is not a fundamental scaling relation.} 

We are also able to rule out bulge-to-total ratio in contributing to the SFR of star-forming galaxies, hence morphological parameters appear to have no significant role. This confirms that there is no secondary dependence of the SK relation on morphology. We again reiterate that in this work we are only considering star-forming galaxies and that with quiescent systems included B/T may influence SFR \citep[][]{Bluck_bulge_2014MNRAS.441..599B}.

\section{Results: High-redshift galaxies}

\label{sec:high-z}

\begin{figure}
    \centering
    \includegraphics[width=\columnwidth]{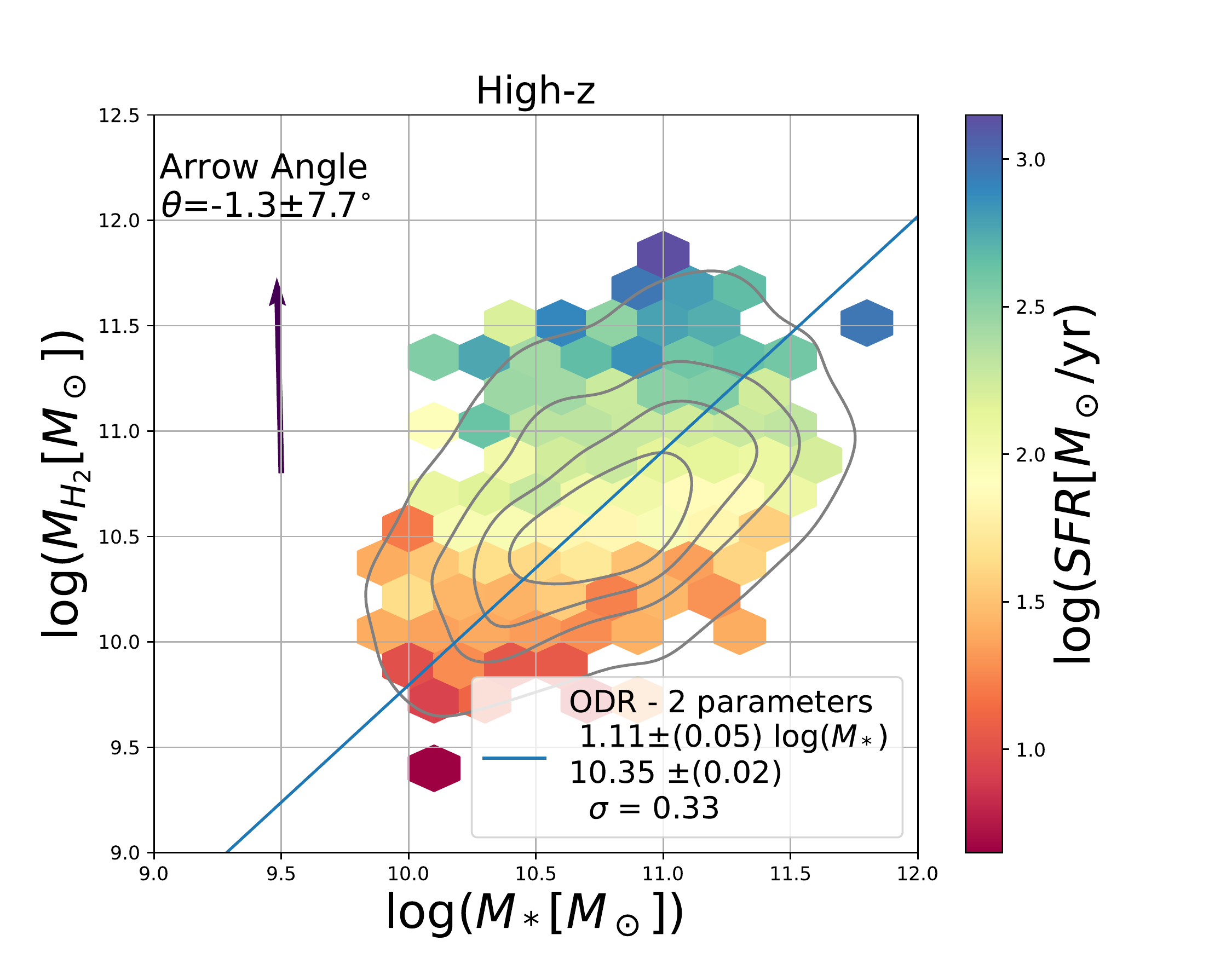}
    \caption{2D histogram  of the molecular gas mass vs stellar mass colour coded by the mean log star formation rate for high-redshift galaxies (z>0.5). Contours show the distribution of galaxies with the
    outer density contour contains 90\% of the galaxies and the blue line is an orthogonal distance regression fit to the data. The partial correlation coefficient arrow points in the direction of the greatest increasing gradient for the SFR (color-coded quantity). The gradient arrow shows that the star formation rate is almost totally determined by the molecular gas mass with little to no contribution on the stellar mass. The ODR line reveals the existence of the MGMS. However, this diagram is affected by the redshift evolution of molecular gas at fixed stellar mass, which broadens the distribution. This issue is tackled by the analysis in redshift bins later on.}
    \label{fig:high_z_main}
\end{figure}

\begin{figure}
    \centering
    \includegraphics[width=\columnwidth]{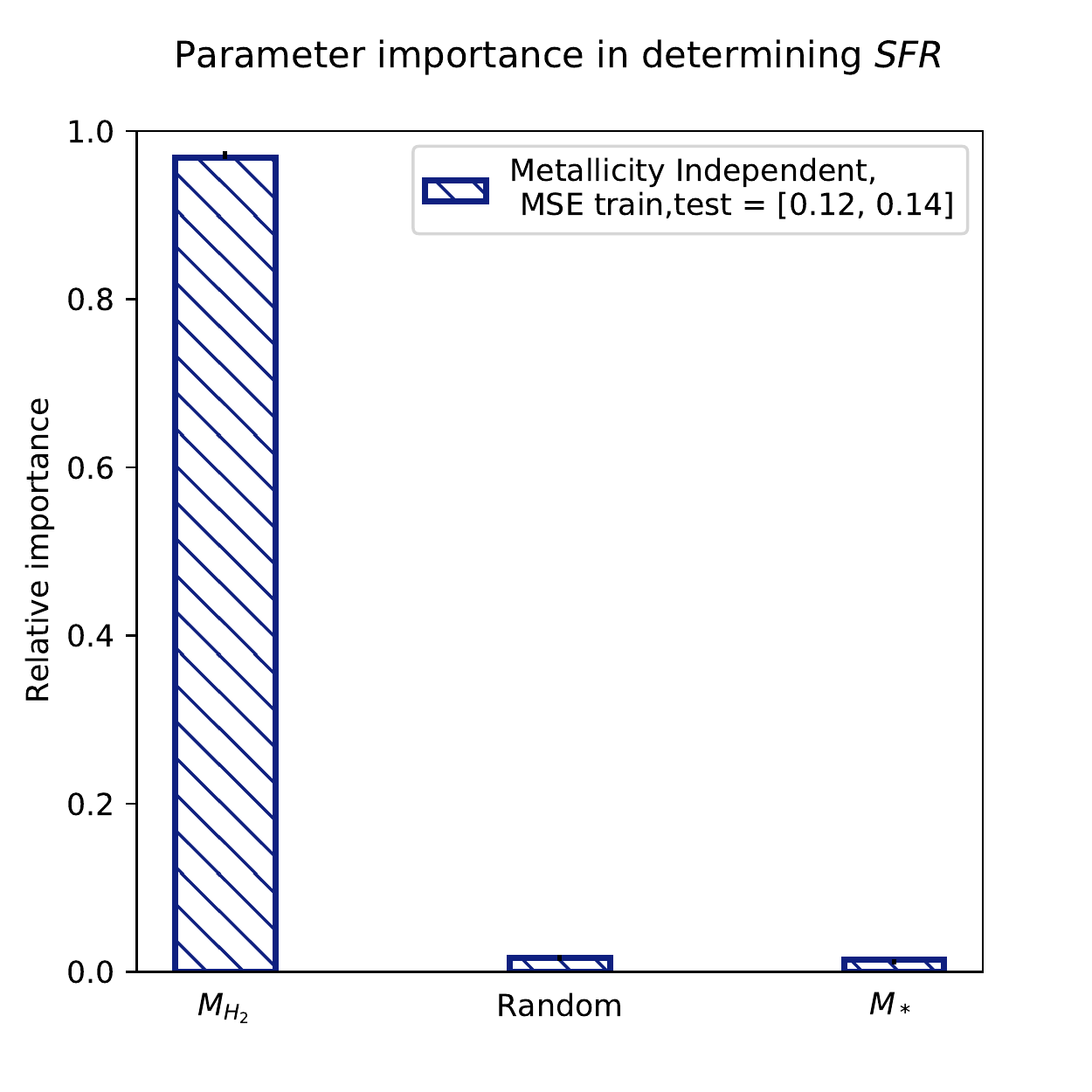}
    \caption{Relative importance of the molecular gas mass ($M_{H_2}$), stellar mass ($M_*$) and a (control) uniform random variable (Random), in determining the SFR, for high-redshift galaxies, according to the Random Forest regression analysis. The Mean Squared Error (MSE) is reported for the training and test samples, showing that the model is not overfit. The errors are obtained by bootstrap random sampling 100 times.
    The bar-chart shows that, also at high redshift, the SFR is almost entirely determined by the molecular gas mass with the contribution from stellar mass being as unimportant as that of a random variable. This confirms that at high-redshift the SFMS is not a fundamental scaling relation.}
    \label{fig:high_z_RF}
\end{figure}

The next stage is to explore whether there are differences between the scaling relations of galaxy populations in the local universe and those at higher redshift. We compare whether M$_{H_2}$ or M$_*$ is the primary driver of SFR at high-z.

We start by utilising the whole sample of z>0.5 galaxies from PHIBBS (824 galaxies). This includes galaxies where molecular gas mass was obtained via CO measurements and also galaxies where it was obtained via dust masses.

Figure \ref{fig:high_z_main} shows a 2D histogram of molecular gas mass versus stellar mass colour coded by the mean log star-formation rate in each bin. It is the high-z analogue of Figure \ref{fig:Mstar-H2-SFR-Starforming}. The density contours enclose 90\% of the galaxies. The blue line is an orthogonal distance regression best fit to the data.
The arrow points in the direction of the greatest increasing gradient of the colour-coded quantity (SFR).
The arrow angle, with its error, is consistent with zero degrees meaning that, as it was found for the local galaxies, the star formation rate is almost entirely determined by the molecular gas mass with little to no direct dependence on the stellar mass, as also visualised by the colour gradient.
However, in this case the observed effect may be a consequence of the fact that the MGMS could evolve with redshift (hence $M_{H2}$ increase with redshift at a given stellar mass) and we are mixing up all the redshifts (i.e. a combination of redshifts ranging from z$\sim$0.5 up to z$\sim$4). This will be addressed in the next section.

Figure \ref{fig:high_z_RF} is a bar-chart showing the parameter importances of the molecular gas mass ($M_{H_2}$), a uniform random variable (Random) and the stellar mass ($M_*$) in determining the star formation rate according to the Random Forest regression analysis. It is clear that the molecular gas mass is by far the most important parameter in determining the star formation rate whilst the stellar mass has no predictive power. This reveals that the Schmidt Kennicutt gas fuelling relation remains much more important in determining the SFR than the SFMS at high redshift.
Hence we can conclude that the SFMS is not a fundamental scaling relation relation at high-z, but simply a by-product of the SK relation, through the MGMS.

\subsection{The redshift evolution of the MGMS}
\label{sec:z_bins}

\begin{figure*}
    \centering
    \includegraphics[width=2.0\columnwidth]{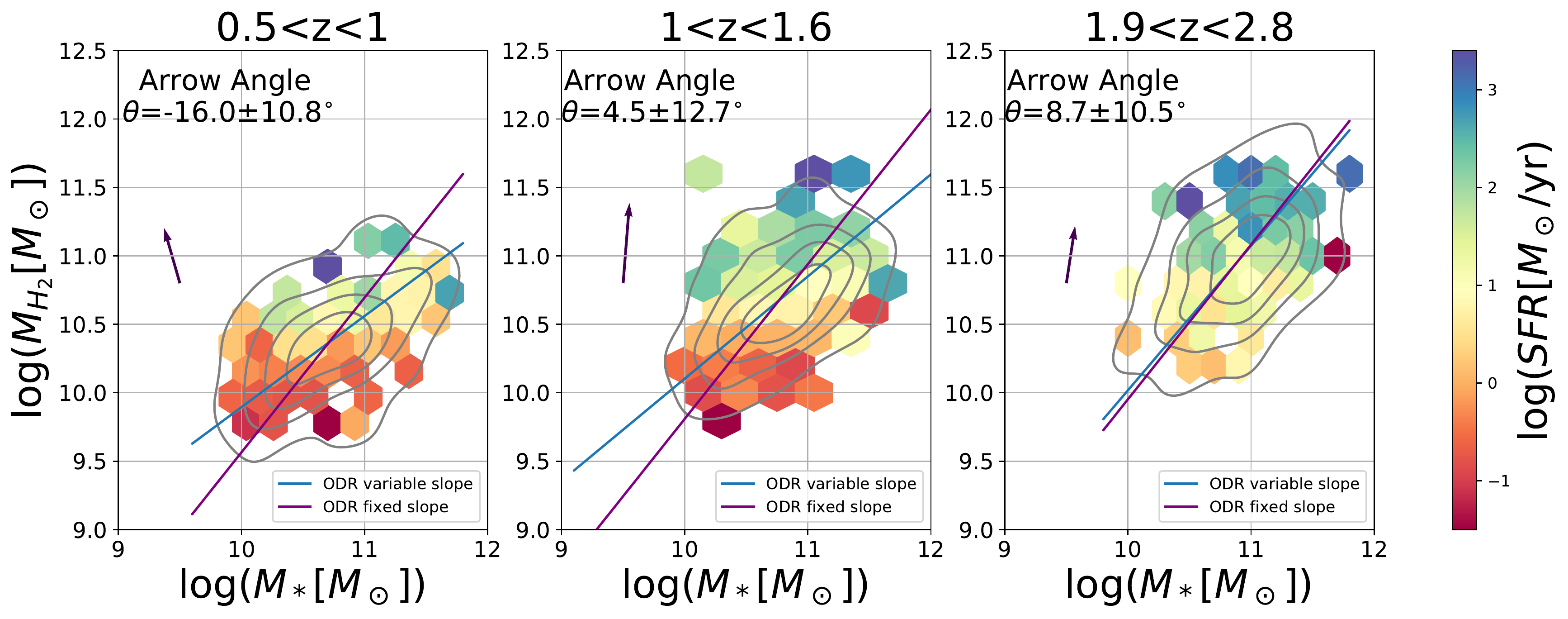}
    \caption{2D hexagonal binned plot of molecular gas mass vs stellar mass colour coded by mean log star formation rate in each bin. Each panel corresponds to a different redshift bin, as indicated on the top of each panel. Contours show the distribution of galaxies in each redshift bin, with the outer density contour enclosing 90\% of the data points.  The ODR line is calculated with a varying slope (blue) and with the slope fixed to the value from the local universe sample (purple). The ODR lines show the existence of MGMS at different redshifts.
    Moreover, at a given stellar mass, the molecular gas increases with redshift, revealing that the MGMS evolves with redshift.
    The arrow points in the direction of largest increasing gradient in SFR. Both the color shading and the gradient arrow shows that in each redshift bin the SFR is almost entirely driven by molecular gas mass with an almost negligible contribution from stellar mass, illustrating that the SFMS is not a fundamental relation also at high redshift.}
    \label{fig:binned_z_scatter}
\end{figure*}

\begin{figure}
    \centering
    \includegraphics[width=1\columnwidth]{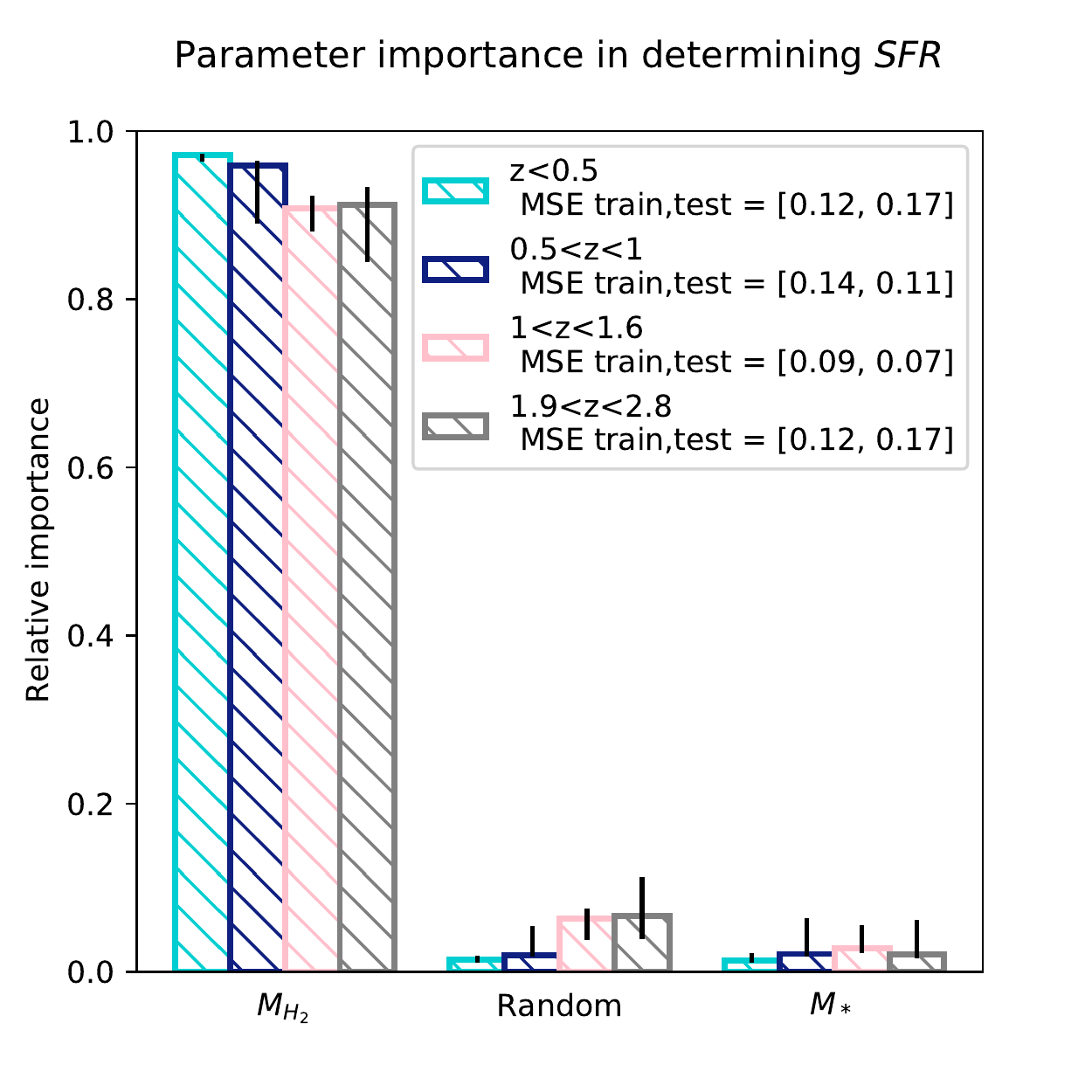}
    \caption{Bar-charts for each redshift bin, showing the parameter importance of the molecular gas mass ($M_{H_2}$), a (control) uniform random variable (Random) and the stellar mass ($M_*$) in determining to the star formation rate in different redshift bins. The Mean Squared Error (MSE) is again reported for the training and test samples to show that the model is not overfit or underfit. Errors are obtained by bootstrap random sampling 100 times.
    Once again, this shows that, for each redshift bin,  the SFR is entirely determined by the molecular gas mass with no contribution from the stellar mass, i.e. the stellar mass is as unimportant as the uniform random variable. }
    \label{fig:RF_binned}
\end{figure}

The next stage is to split the PHIBBS sample into separate bins of redshift so that we can explore any evolution of the scaling relations with redshift.

The redshift distribution of the galaxies and an explanation of our binning scheme is given in the Appendix (Figure \ref{fig:redshifts}). We split the galaxies into three bins of redshift: 0.5<z<1, 1<z<1.6 and 1.9<z<2.8, enabling us to include the majority of the sample in the analysis. 
The number of galaxies in each bin are as follows: the 0.5<z<1 bin has 406 galaxies; the 1<z<1.6 bin, 354 galaxies; and the 1.9<z<2.8 bin, 182 galaxies. 

Figure \ref{fig:binned_z_scatter} is a color-coded 2D hexagonal binned plot similar to Figure \ref{fig:high_z_main} but split into three separate panels, each corresponding to one redshift bin. As before, the density contours show the distribution of galaxies, with the outer enclosing 90\% of the galaxies.  The best-fit line is calculated via orthogonal distance regression (ODR). 
The partial correlation coefficient arrow points in the direction of the greatest increase in star formation rate.
The partial correlation coefficient arrow angles and the colour gradient immediately reveal that the star formation rate depends almost entirely on the molecular gas mass with little to no dependence on the stellar mass in each individual redshift bin. This also confirms that the previous results, where we combined all the redshifts into one bin, were not being skewed by any particular redshift range and not an artifact (or not only) of merging all redshifts together. 

Figure \ref{fig:RF_binned} shows the random forest results (as in Figure \ref{fig:high_z_RF}) for each redshift bin (including the PHIBBS local sample). What is immediately apparent is that, again, the SFR depends primarily on the molecular gas mass {\it at any redshift}.  The plot shows that, at all redshifts considered, the Schmidt-Kennicutt relation is far more important than the star forming main sequence. It also again confirms that the previous random forest result was not biased by any particular redshift range.

In addition, as it was the case for the local universe, we are able to define a Molecular Gas Main Sequence relation for each redshift bin using ODR. We present two approaches here, the first allows the slope of the ODR line to vary and the second fixes it at the value obtained for the local galaxies. The first approach has the advantage of including a possible evolution in the MGMS slope, but the second approach is better able to deal with the limited sample sizes in each redshift bin, as it will be less skewed by noisy data. We use again the MGMS linear fit in log space with the form given by Eq.\ref{eq:odr_general}.

\begin{table*}
\caption{Molecular Gas Main Sequence (MGMS) fits with orthogonal distance regression (ODR) for different bins of redshift. The ODR lines themselves are shown in Figure \ref{fig:binned_z_scatter}. m is the gradient of the ODR line for the varying gradient case while $\rm log(M_{H2})_{10.5}$ is the value of $\rm log(M_{H_2})$ at $\rm log(M_*/M_\odot=10.5)$.  $\rm log(M_{H2})_{10.5|1.13}$ is the value of $\rm log(M_{H_2})$ at $\rm log(M_*/M_\odot=10.5)$ for the case where m=1.13, i.e. we have fixed the gradient of the ODR line to that of the local universe. This table shows that, as the redshift increases, the molecular gas mass at a given stellar mass also increases, hence there is a clear evolution of the MGMS relation with redshift. $\sigma$ is the scatter of the points from the ODR best-fit - we see that it is of the order 0.3dex for each redshift bin.}
\centering
\label{table:fits}
\begin{tabular}{l c c c c c}
\toprule
\multirow{2}{*}{Redshift bin}  &   \multirow{2}{*}{\# of galaxies} &   \multirow{2}{*}{m} & \multirow{2}{*}{$\rm log(M_{H2})_{10.5}$} & \multirow{2}{*}{$\rm log(M_{H2})_{10.5|1.13}$} & \multirow{2}{*}{$\sigma$} \\[+5pt]

\midrule

0.0 < z < 0.1 & 409 & 1.13 & 9.42 & 9.42 & 0.26\\

0.5 < z < 1.0 & 255 & 0.67 & 10.21 & 10.10 & 0.31 \\

1.0 < z < 1.6 & 354 & 0.75 &  10.46 & 10.35 & 0.30 \\

1.9 < z < 2.8 & 182  & 1.06 & 10.52 & 10.49  & 0.30 \\

\bottomrule
\end{tabular}
\end{table*}

Table \ref{table:fits} contains the values of the coefficients of the fits for a local universe sample (the PHIBBS galaxies with z<0.1) and the three bins of redshift. For the redshift bins, m and $\rm log(M_{H2})_{10.5}$ are the values for the case where the gradient (m) is allowed to vary whilst $\rm log(M_{H2})_{10.5|1.13}$ is the value for the case that the gradient m is fixed to that of the local galaxies in PHIBBS (m=1.13). The ODR lines themselves are plotted in Figure \ref{fig:binned_z_scatter}. We chose to use the local galaxies from the PHIBBS sample as a comparison rather than our own selected sample in order to minimise possible biases between the samples (for example our own sample only contains star-forming galaxies).

From Table \ref{table:fits} and Figure \ref{fig:binned_z_scatter} it is clear that, as the redshift increases, the ODR fit is offset to higher values of $\rm log(M_{H_2})$ for a fixed $M_*$, i.e. $\rm log(M_{H2})_{10.5}$ and $\rm log(M_{H2})_{10.5|1.13}$ increase with redshift. In other words, the MGMS evolves with redshift. We also note that the scatter around the MGMS remains at $\sigma\sim$0.30dex or less regardless of the redshift bin. This suggests, as far as we can say with our limited statistics, the MGMS remains an intrinsically tight relation also at high redshift.

We quantitatively estimate how the MGMS would evolve by assuming the constant gradient scenario and quantifying the rate of increase in the constant $\rm log(M_{H2})_{10.5|1.13}$. To do this we re-bin the data into finer bins. We then calculate the value of $\rm log(M_{H2})_{10.5|1.13}$ per bin. Our minimum redshift bin is z=0.2.
Subsequently, we fit a simple second order polynomial curve to the values of the intercepts for each bin, obtaining a relation of the form
\begin{equation}
    \rm  log(M_{H2})_{10.5} = -0.17\, (\pm 0.03)\,z^2 + 0.84 (\pm 0.09)\,z + 9.77\, (\pm 0.06)\
\label{eq:MGMS_intercept}
\end{equation}
which describes the redshift evolution of the intercept and illustrated in the top-left panel of Fig.\ref{fig:SFMS_evo}.
The intercept increases as the redshift increases, until it reaches a turning point at   z=2.3, i.e. around ``cosmic noon''. 

We note that the extrapolation to z=0 does not match the z=0 bin. This may be due to a possible bias in the high-z sample for which only detections are reported (we have seen that this is not an issue for the local sample, but could be a problem at high-z). This prompts for a more extensive analysis in which biases and upper limits are explored in the PHIBBS sample, but such analysis is beyond the scope of this paper.

By using Eq.\ref{eq:MGMS_intercept} we can obtain the equation deriving the redshift evolution of the MGMS as

\begin{align}
    \rm log(M_{H_2})= 1.13\, [log(M_*)-10.5]\, -0.17\, (\pm 0.03)\,z^2 \nonumber \\+ 0.84 (\pm 0.09)\,z + 9.77\, (\pm 0.06)\
    \label{eq:MGMS}
\end{align}

\subsection{Evolution in the SFMS is a consequence of evolution in SK \& MGMS}

\begin{figure*}
    \centering
    \includegraphics[width=1\columnwidth]{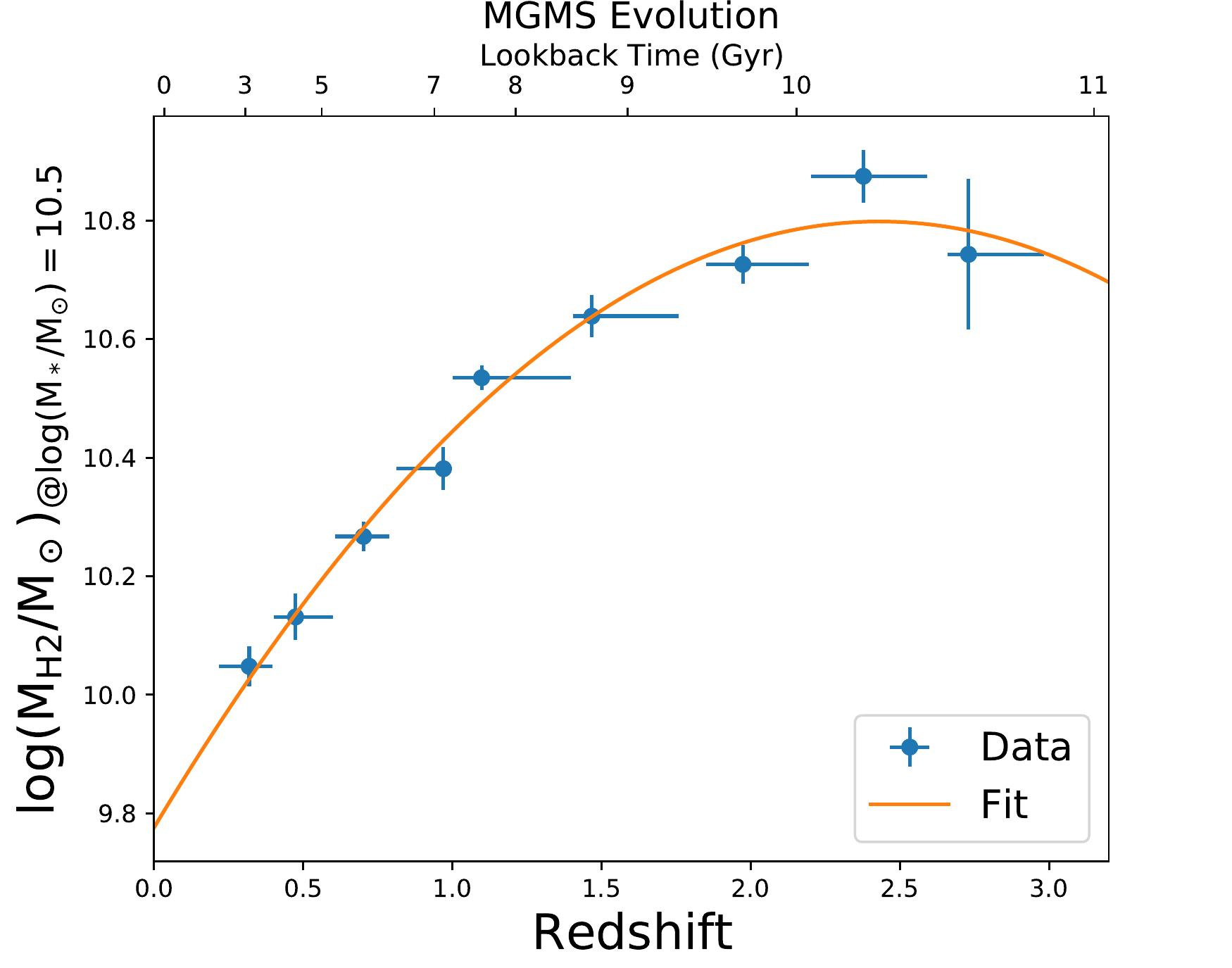}
    \includegraphics[width=1\columnwidth]{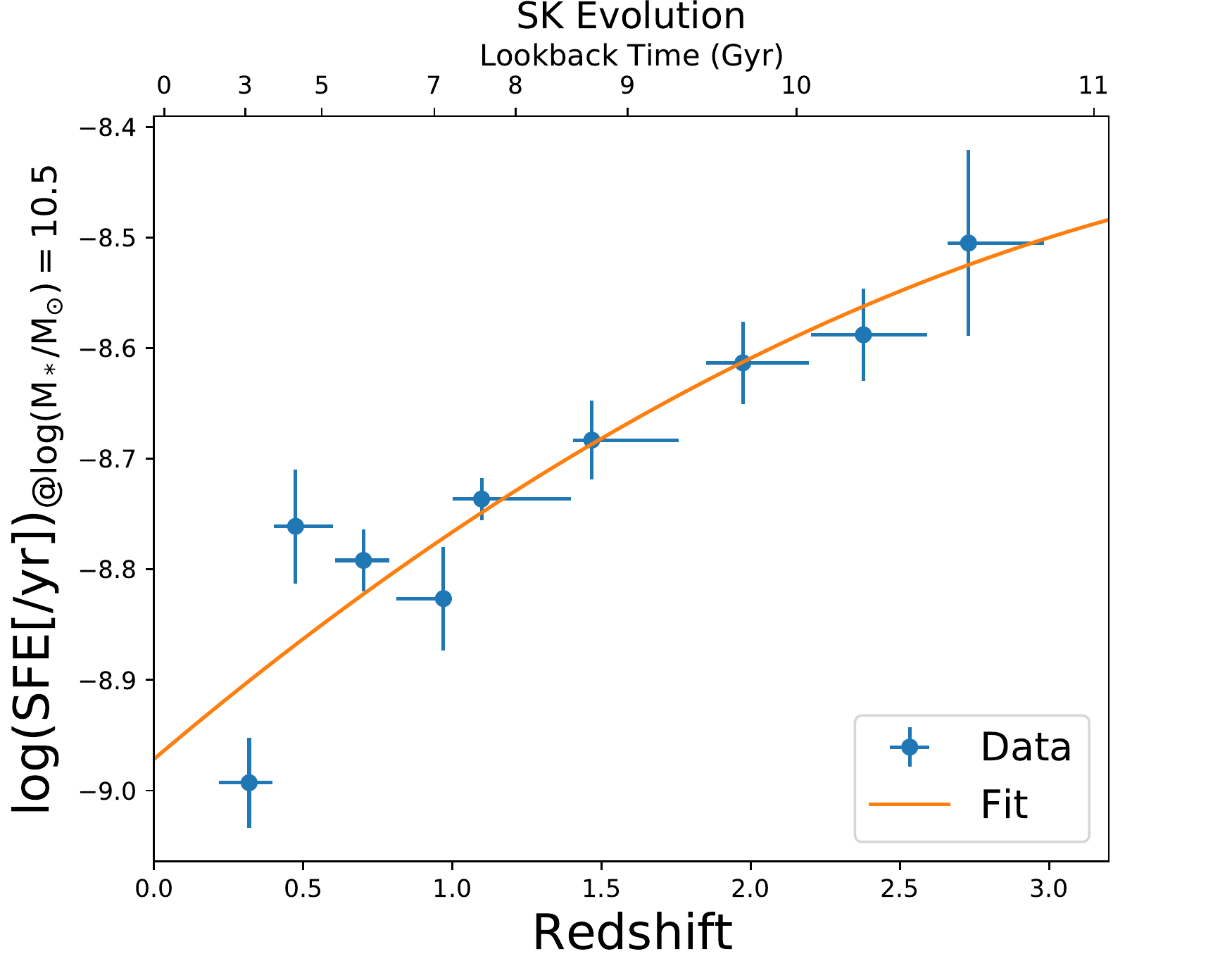}
    \includegraphics[width=1.1\columnwidth]{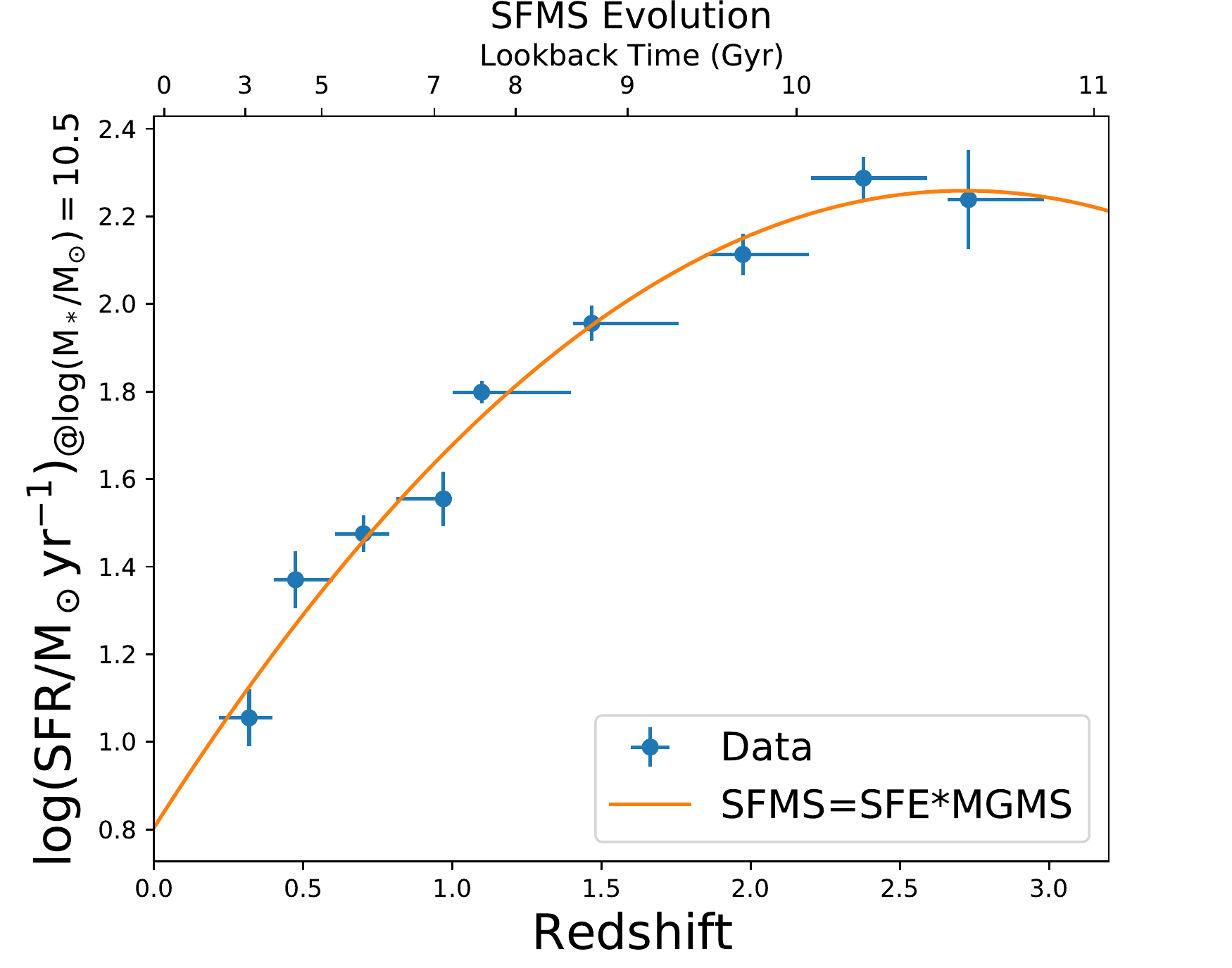}
    \includegraphics[width=0.8\columnwidth]{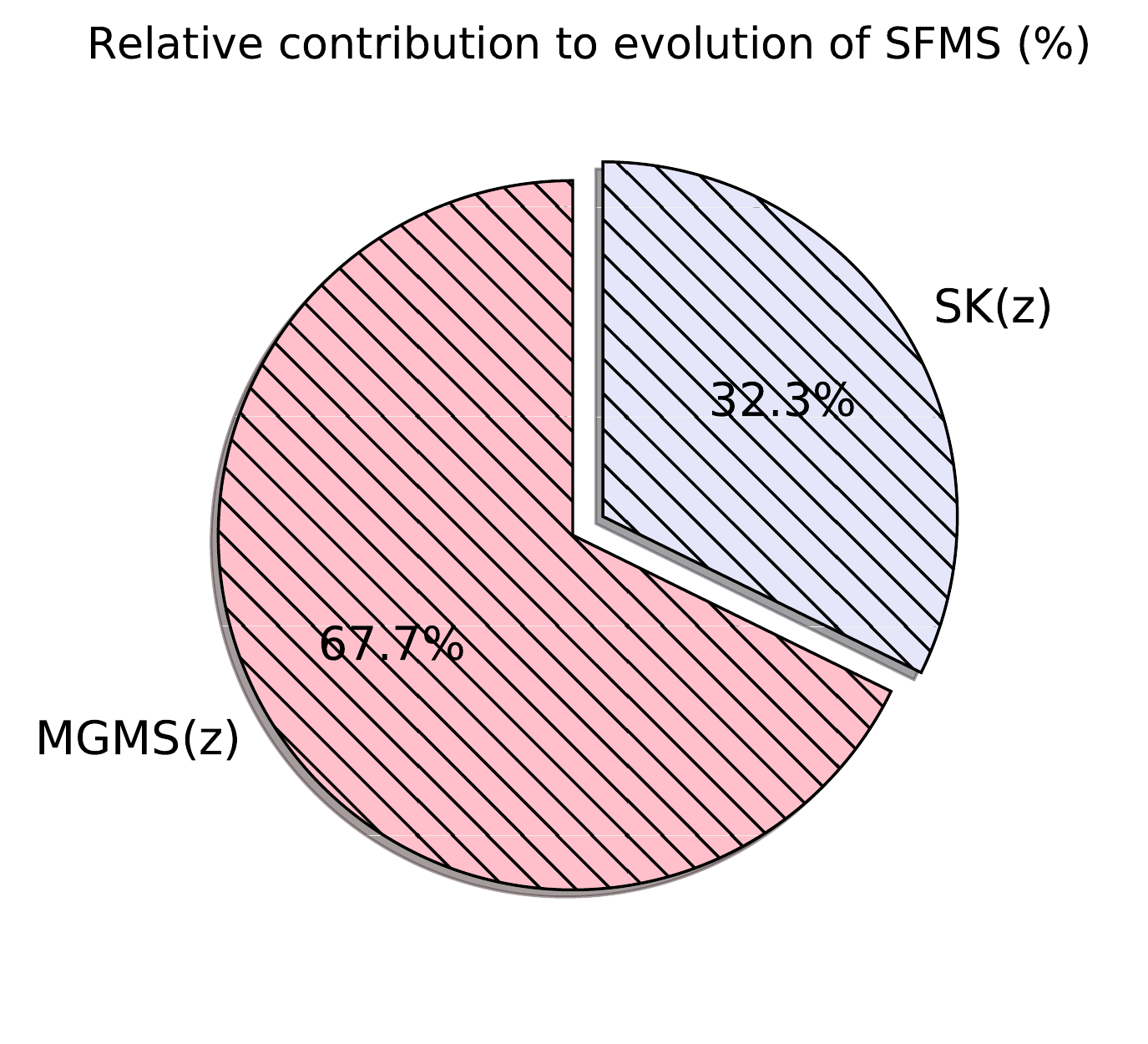}
    \caption{The redshift evolution of the Molecular Gas Main Sequence (MGMS, upper left), Star Formation Efficiency (SFE, upper right), Star Forming Main Sequence (SFMS, lower left), and the relative contributions of the MGMS and SFE in driving the observed SFMS evolution (bottom-right). 
    $\rm log(M_{H2})_{@log(M_*/M_{\odot})=10.5}$ is the value of log(M$_{H2}$/M$_\odot$) at log(M$_\star$/M$_\odot$)=10.5 for the fixed gradient best fit MGMS line case, whilst $\rm log(SFR)_{@log(M_*/M_{\odot})=10.5}$ and $\rm log(SFE)_{@log(M_*/M_{\odot})=10.5}$ are the SFR and SFE respectively at the same fixed stellar mass. The error bars in the redshift axis are the width of each bin. The y axis errorbars are the standard error. The orange line in the top two panel is a simple 2$^{nd}$ order fit to the data, while in the bottom left panel the orange line is {\it not} a fit to the data, it is simply tracing the equation resulting from the combination of the fits in the top two panels. The latter shows that the observed SFMS evolution with redshift can be obtained by simply combining the redshift evolutions of the MGMS and the SFE  via SFMS(z)= SFE(z) $\times$ MGMS(z). The lower left plot shows that this combination provides an almost perfect fit to the points of $\rm log(SFR)_{@log(M_*/M_{\odot})=10.5}$. This means that the evolution of the SFMS with redshift can be determined by the evolution of the SK and MGMS relations. The lower right plot reveals the relative contributions of the redshift evolution of the MGMS and SFE to the SFMS. The size of the segments show that primarily it is the MGMS that drives the SFMS's evolution with redshift, whilst the the SFE (SK) plays a secondary role. }
    \label{fig:SFMS_evo}
\end{figure*}

In the previous section we defined a parameterisation of the redshift evolution of the MGMS. In this section we explore the redshift evoluton of the SK relation (i.e. the potential evolution of the $\rm SFE=SFR/M_{H_2}$) and explore whether the redshift evolution of the SFMS can be simply explained as a consequence of the evolution of the MGMS and SK. 

We start by determining the evolution of the SK. We again use the same bins of redshift to bin the star formation efficiency (SFE). We fit the SFE as function of stellar mass (which results into a nearly flat dependence) and determine its intercept at a constant stellar mass, i.e. $\rm @log(M_*/M_{\odot})=10.5$. We then fit a 2nd order polynomial to the redshift bins.
The resulting parametrisation of the evolution of the SFE is given by

\begin{equation}
    \rm log(SFE)= -0.024\,(\pm 0.04)\, z^2 + 0.23\,(\pm 0.12)\, z -8.97 (\pm 0.08)\
    \label{eq:SFE_intercept}
\end{equation}

The resulting evolution is shown in the top-right panel of Figure \ref{fig:SFMS_evo}.

Finally, we explore the evolution of the SFMS, i.e. the evolution of the SFR at the same given reference stellar mass as for the MGMS and for the SFE. Specifically, we determine the SFMS in the same redshift bins, and fit the data with a log-linear equation, and derive the SFR intercept again at $\rm @log(M_*/M_{\odot})=10.5$. The resulting values as a function of redshift are shown in blue in the bottom-left panel of Figure \ref{fig:SFMS_evo}.

We do {\it not} fit the observed evolution as in the previous two cases. Instead,
we make the working hypothesis that the redshift evolution of the SFMS is simply a consequence of the evolution of the MGMS and of the SFE, i.e.

\begin{equation}
   \rm  SFMS(z) = SFE(z) \times MGMS(z)
   \label{eq:model}
\end{equation}

and derive the evolution of the SFMS(z) {\it inferred} by replacing the SFE(z) and MGMS(z) from Eq. \ref{eq:MGMS_intercept} and Eq. \ref{eq:SFE_intercept} into Eq.\ref{eq:model}. The resulting, {\it expected} evolution of the SFMS is shown with the orange line in the bottom-left diagram of Figure \ref{fig:SFMS_evo}, which nicely passes through the observational data. In other words, the orange curve in the SFMS(z) bottom-left panel of Figure \ref{fig:SFMS_evo} is not a fit to the data, is simply derived by combining the analytical expressions (orange lines) of the MGMS(z) top-left and SFE(z) top-right panels. This suggests that the redshift evolution of the SFMS is driven by the evolution of the MGMS and the evolution of the SFE (i.e. SK relation). 

Figure \ref{fig:SFMS_evo} also shows that it is primarily the redshift evolution of the MGMS that drives the SFMS with the SFE being a secondary driver. This is illustrated more quantitatively by the bottom-right panel of Figure \ref{fig:SFMS_evo}, which gives the relative contribution of MGMS(z) and SFE(z) to the evolution of the SFMS(z). 

From this we can conclude that, not only is the SFMS not a fundamental scaling relation, but that its observed evolution with redshift appears to be entirely driven by the MGMS and the SK relation, i.e. the two relations which cause the SFMS are also responsible for its redshift evolution.

\subsection{Fuel or Efficiency?}

\begin{figure}
    \centering
    \includegraphics[width=\columnwidth]{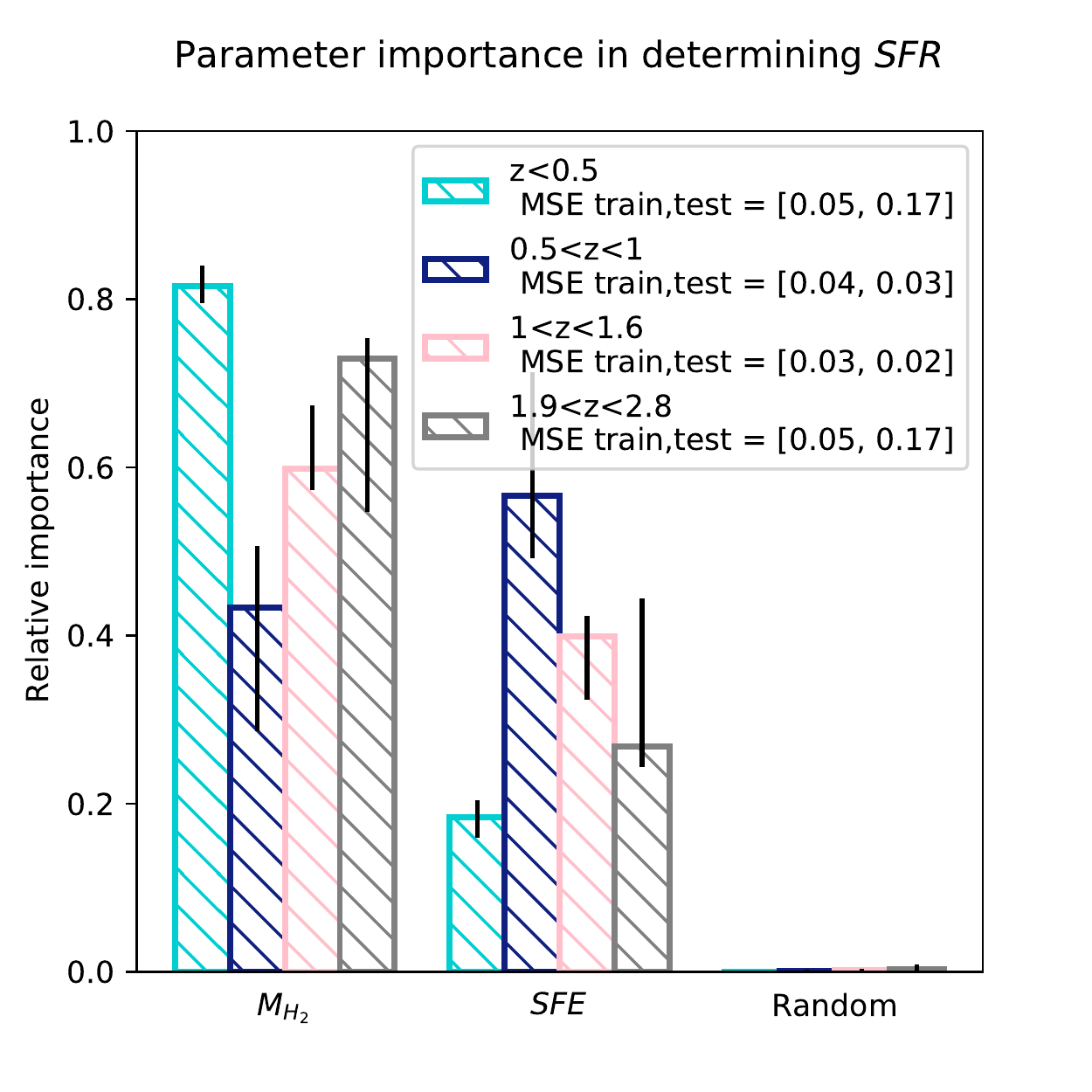}
    \caption{Is star formation fuel or efficiency driven at each redshift epoch? This figure shows the parameter importances in determining the star formation rate. The parameters included are molecular gas mass, star formation efficiency and a uniform random variable. This enables a test of whether the key driver of SFR is fuel or a combination of everything else (e.g. magnetic fields, turbulence etc.).  The different coloured bars correspond to the redshift bins. The errors are obtained by bootstrap random sampling 100 times. This figure reveals which of molecular gas mass or star formation efficiency is more important for determining the star formation rate at different redshifts, i.e. is it the amount of fuel for star formation or the efficiency at which it is converted into stars that is the key factor. 
    The bars show that each epoch the amount of molecular gas and the SFE both contribute to determining the SFR.}
    \label{fig:RF_binned_sfe}
\end{figure}

In addition to the previous analysis, we are able to investigate whether star formation is primarily driven by amount of gas or efficiency of conversion of the gas into stars for each of these redshift bins.
To do this we again utilise the random forest, only this time including the star formation efficiency (SFE=SFR/$M_{H_2}$). 
We can consider SFE as a combination of all the other unknowns that contribute to star formation, aside from simply the amount of fuel. For example, these unknowns could be turbulence, magnetic fields, gas dispersion, metallicity and further quantities.
Figure \ref{fig:RF_binned_sfe} is a bar-chart showing the results of this random forest regression for each redshift bin (including the local sample). The plot shows how at each redshift the relative ratio between the importances of molecular gas mass and SFE varies. We see that in the local universe the dominant driver is the amount of fuel whilst at higher redshifts SFE appears to play a stronger role although with the amount of fuel still being slightly more important. The same importance of the SFE was found by \citet{Saintonge2022arXiv220200690S}, \cite{Tacconi2020}, \cite{Piotrowska2019}, and \cite{Piotrowska2022MNRAS.512.1052P}.

A possible issue regarding this result is the measurement uncertainty and, specifically, the fact that SFE is ultimately derived from SFR hence their uncertainties are by construction linked. The measurement uncertainty on SFE also contains a contribution from $M_{H_2}$ so this could artificially increase the importance of SFE compared to $M_{H_2}$.

\section{AGN}

\label{sec:AGN}

\begin{figure}
    \centering
    \includegraphics[width=\columnwidth]{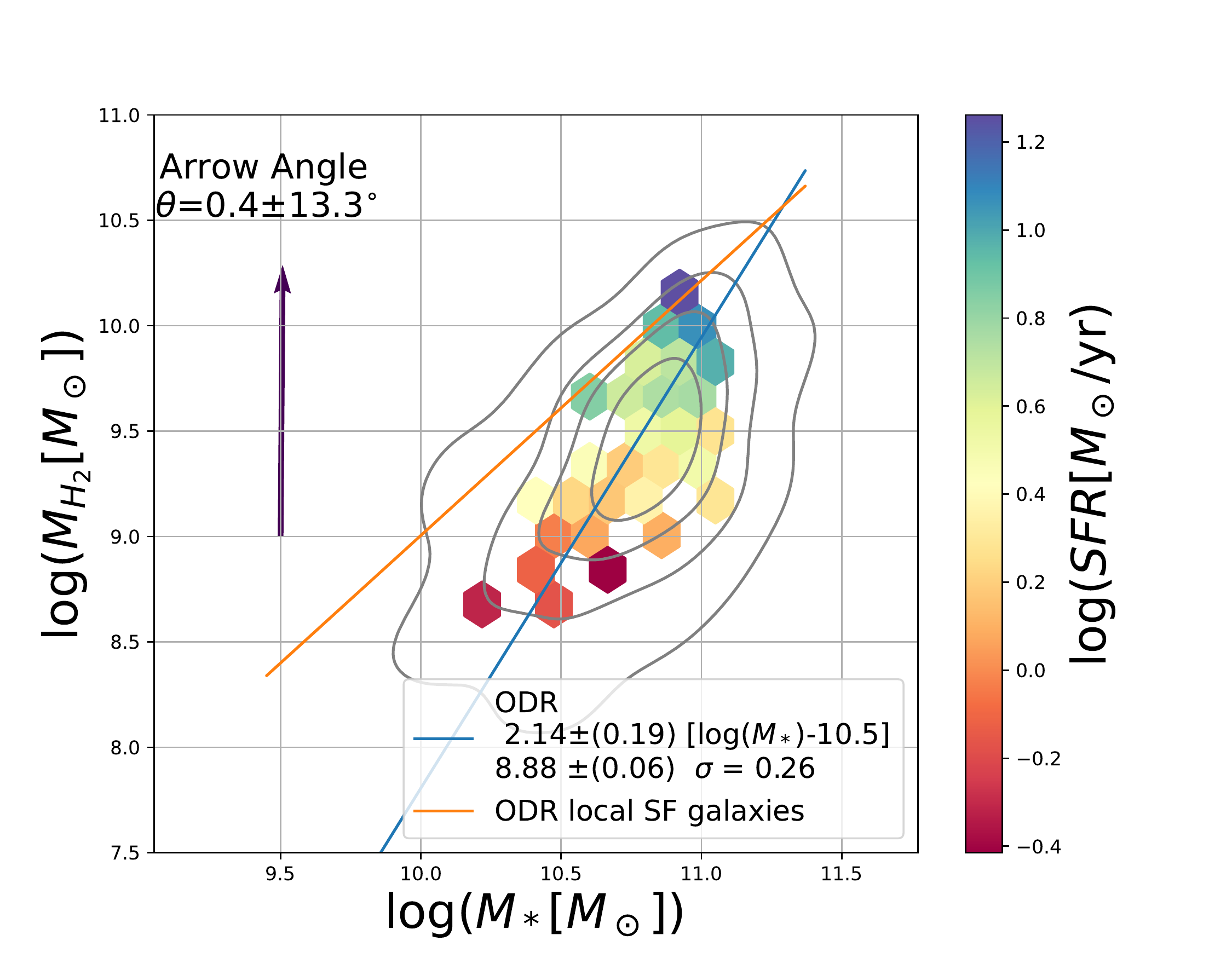}
    \caption{Molecular gas mass versus stellar mass colour coded by star formation rate for the BAT sample AGN. The bin value is the mean log star formation rate of all the AGN in that bin. The blue line is calculated via orthogonal distance regression which minimises scatter from the line for both stellar mass and molecular gas mass. The contours reveal the density distribution of galaxies with the outer contour enclosing 90\% of the galaxy population. The partial correlation coefficient arrow points in the direction of greatest increasing gradient in star formation rate. We see that the star formation rate is driven completely by the molecular gas mass with no significant contribution from the stellar mass. The figure shows that for AGN host galaxies the star forming main sequence is unimportant compared to the Schmidt-Kennicutt relation in determining the star formation rate. The blue  ODR line shows a MGMS for AGN host galaxies with a scatter of $\sigma=0.26$dex. The orange ODR line shows the best-fit MGMS for the local SF galaxies.}
    \label{fig:AGN-PCC}
\end{figure}

\begin{figure}
    \centering
    \includegraphics[width=\columnwidth]{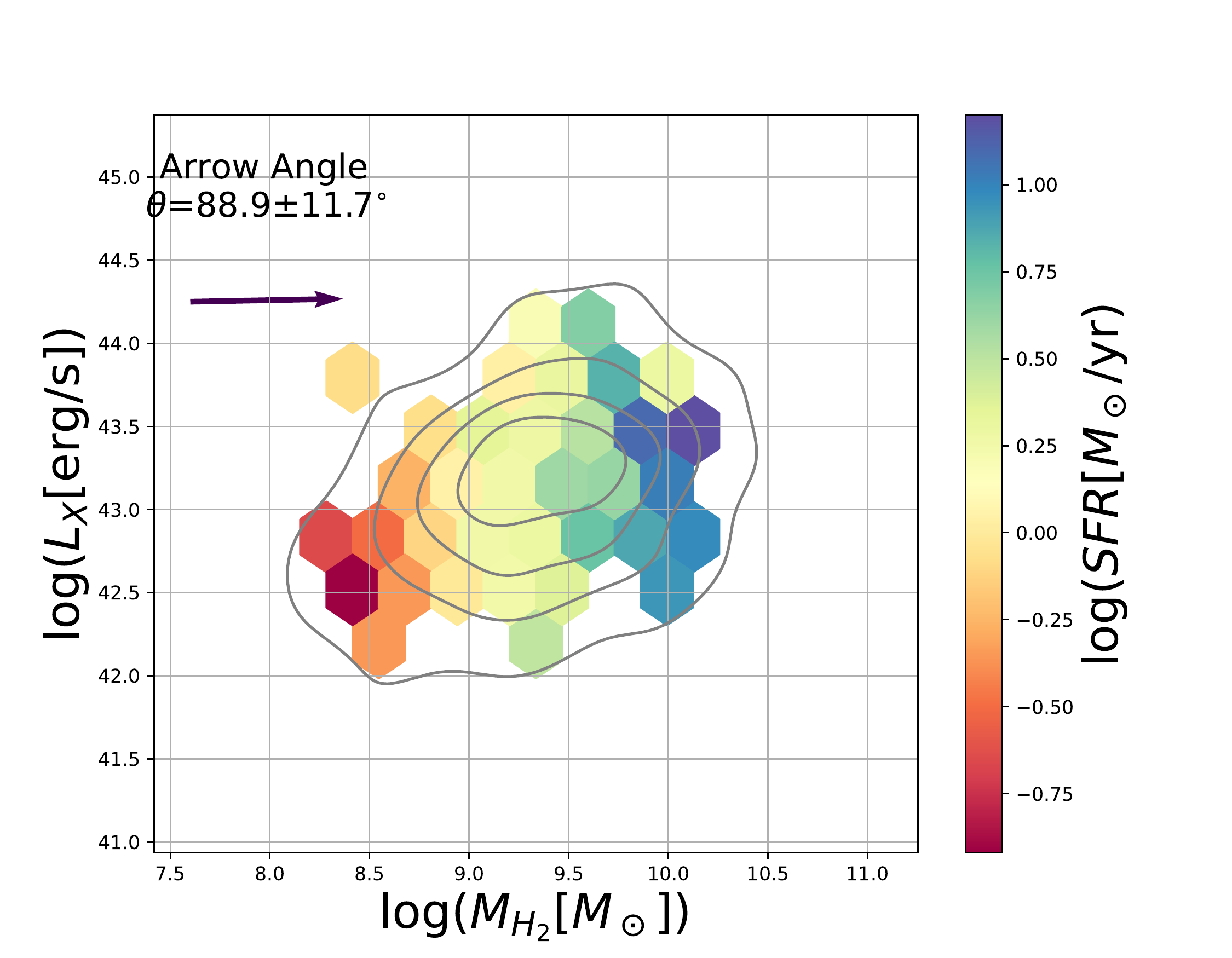}
    \caption{2D hexagonally binned plot showing AGN x-ray luminosity against molecular gas mass, colour coded by the mean log star formation rate in each bin. The partial correlation coefficient gradient arrow points in the direction of greatest increase in star formation rate. The outer density contour encloses 90\% of the galaxies. The arrow reveals that the star formation rate is totally driven by the amount of molecular gas with no contribution from the x-ray luminosity (i.e. instantaneous AGN feedback has little to no effect on star formation rate). }
    \label{fig:pcc_l}
\end{figure}

\begin{figure}
    \centering
    \includegraphics[width=\columnwidth]{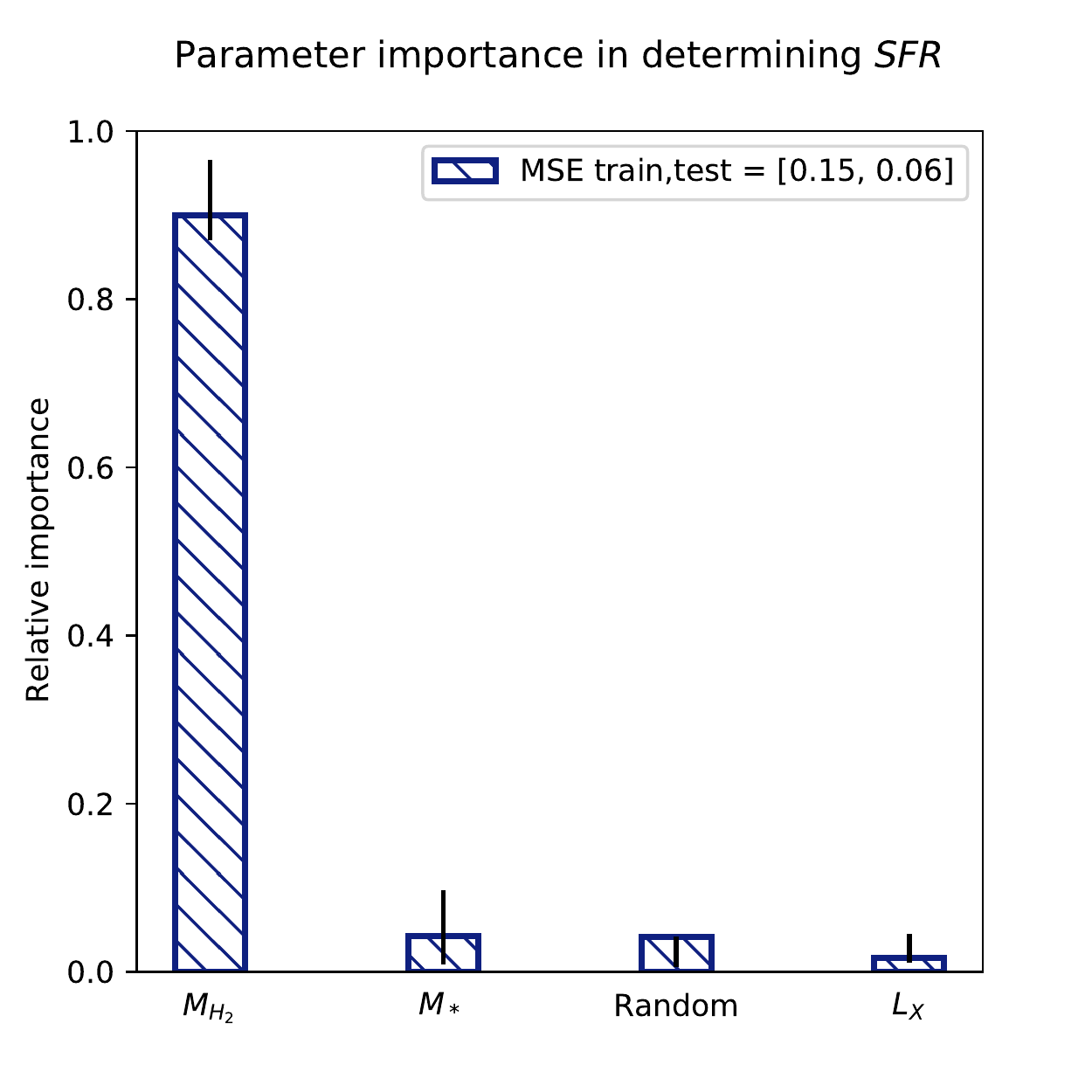}
    \caption{Random forest regression for determining the SFR of the AGN included in the BAT survey. The parameters evaluated are: molecular gas mass (M$_{H_2}$), stellar mass (M$_*$), a uniform random variable, and instantaneous AGN luminosity (L$_X$). The figure shows that, as in the case for the star forming galaxies, the SFR is almost entirely driven by the molecular gas mass, with the contributions from stellar mass being as unimportant as the uniform random variable. This is in agreement with the results of the previous two plots where the partial correlation coefficient gradient arrows show the overwhelming importance of molecular gas mass in determining SFR. It also shows that instantaneous AGN feedback, as traced by $L_X$ plays no significant role.}
    \label{fig:AGN-RF}
\end{figure}

AGN are expected to be a major source of feedback in galaxies by suppressing star formation, either by heating and expelling the ISM (prompt, ejective feedback) and/or by preventing accretion via heating of the CGM (delayed, preventive feedback). It is therefore interesting to explore whether the presence of an AGN affect the scaling relations discussed in this paper.

As we did in previous sections, we can explore which of stellar mass or molecular gas mass is most important in determining the star formation rate, this time for AGN host galaxies. We can also investigate whether AGN luminosity, as traced by hard X-ray luminosity, has an effect on the star formation rate. This will indicate whether the SFR of an AGN host galaxy is strongly impacted by {\it instantaneous} AGN feedback.

Figure \ref{fig:AGN-PCC} shows the 2D histogram of molecular gas mass versus stellar mass, colour coded by SFR (with the colour corresponding to the mean value of the log SFR of the galaxies in that bin) for AGN host galaxies in the BAT-CO sample.
This plot is similar to that in Figure \ref{fig:Mstar-H2-SFR-Starforming}, only this time for the AGN rather than the star-forming galaxies. The partial correlation arrow and the colour gradient clearly show that the SFR almost entirely depends upon the molecular gas mass rather than the stellar mass. As was the case for the star forming galaxies, the arrow angle is consistent with 0$^\circ$.  
This indicates that the Schmidt-Kennicutt relation is far more important for determining the SFR than the SFMS. 

Figure \ref{fig:pcc_l} shows a 2D histogram of molecular gas mass against X-ray AGN luminosity, colour coded again by star formation rate.
The angle of the partial correlation arrow (consistent with zero) and the color gradient clearly illustrates that the SFR is almost completely dependent on the molecular gas mass with no contribution from the AGN luminosity. 
This means that instantaneous AGN feedback does not appear to play any particular role in determining star formation rate, rather SFR is primarily driven by the amount of molecular gas mass.

Figure \ref{fig:AGN-RF} provides the results of the random forest regression in determining the SFR, in the case of the following four parameters: molecular gas mass, stellar mass, the AGN luminosity, and a uniform random variable. As can be seen, the SFR of the AGN hosts is determined almost completely by the molecular gas mass, with the other quantities consistent with the uniform random variable, i.e. not playing a role. This is in complete agreement with the results obtained by the two partial correlation plots (Figures \ref{fig:AGN-PCC} and \ref{fig:pcc_l}). 

 We note that the hexbin plot for these non-BPT selected galaxies looks very different to the BPT-SF selected galaxies in section \ref{sec:localSF}.  This is expected as these AGN host galaxies include a number of quiescent systems, which are on average more massive.

Furthermore, Figure \ref{fig:AGN-PCC} clearly illustrates that AGN host galaxies are also characterised by a MGMS. We can parameterise the MGMS for local AGN using an ODR fit of the form as for normal galaxies (Equation \ref{eq:odr_general}), resulting in
\begin{equation}
    \rm log(M_{H2)}=2.14\,[log(M_*)-10.5]+8.88.
\end{equation}

which is indicated by the blue line in Figure \ref{fig:AGN-PCC}.

We can then compare this AGN-MGMS to the MGMS that we obtained earlier for the local star-forming galaxies (orange line in Figure \ref{fig:AGN-PCC}). If we compare the values of log(M$_{H2}$/M$_\odot$) at log(M$_\star$/M$_\odot$)=10.5 we find that for star forming galaxies log(M$_{H2}$/M$_\odot$)=9.61 whilst for AGN log(M$_{H2}$/M$_\odot$)=8.88. So for this fixed stellar mass, star forming galaxies have on average $\sim$0.7dex more molecular gas mass than AGN host galaxies at this stellar mass. This could indicate that AGN activity can have an effect in reducing the amount of molecular gas either through removal (ejection) or starvation (possibly as a consequence of halo heating). However, the slope of the AGN-MGMS is much steeper than the MGMS for star forming galaxies, hence the two tend to be in agreement at higher masses (where the bulk of the AGN activity happens). Moreover, we warn that the two samples have different selection and observational effects, hence their comparison is not necessarily fair. For instance, our analysis in Sect. \ref{sec:localSF} only included star-forming galaxies (or those selected as such via the [NII]-BPT diagram) whereas AGN hosts also include Green Valley galaxies and quiescent galaxies.

 Regarding the AGN-SK relation, it was already shown by \cite{2021KossBATApJS..252...29K} that AGN follow a Schmidt-Kennicutt relation (as inferred also by our 2D histograms in Fig. \ref{fig:AGN-PCC}, PCC and RF analysis, Fig. \ref{fig:AGN-RF}, as discussed above) and that it is consistent with the Schmidt-Kennicutt relation of non-AGN galaxies; so we do not show that comparison in this paper.

\section{Comparison with previous results}

\subsection{Determining SFE with $\Delta$MS, stellar mass and redshift}

\begin{figure}
    \centering
    \includegraphics[width=\columnwidth]{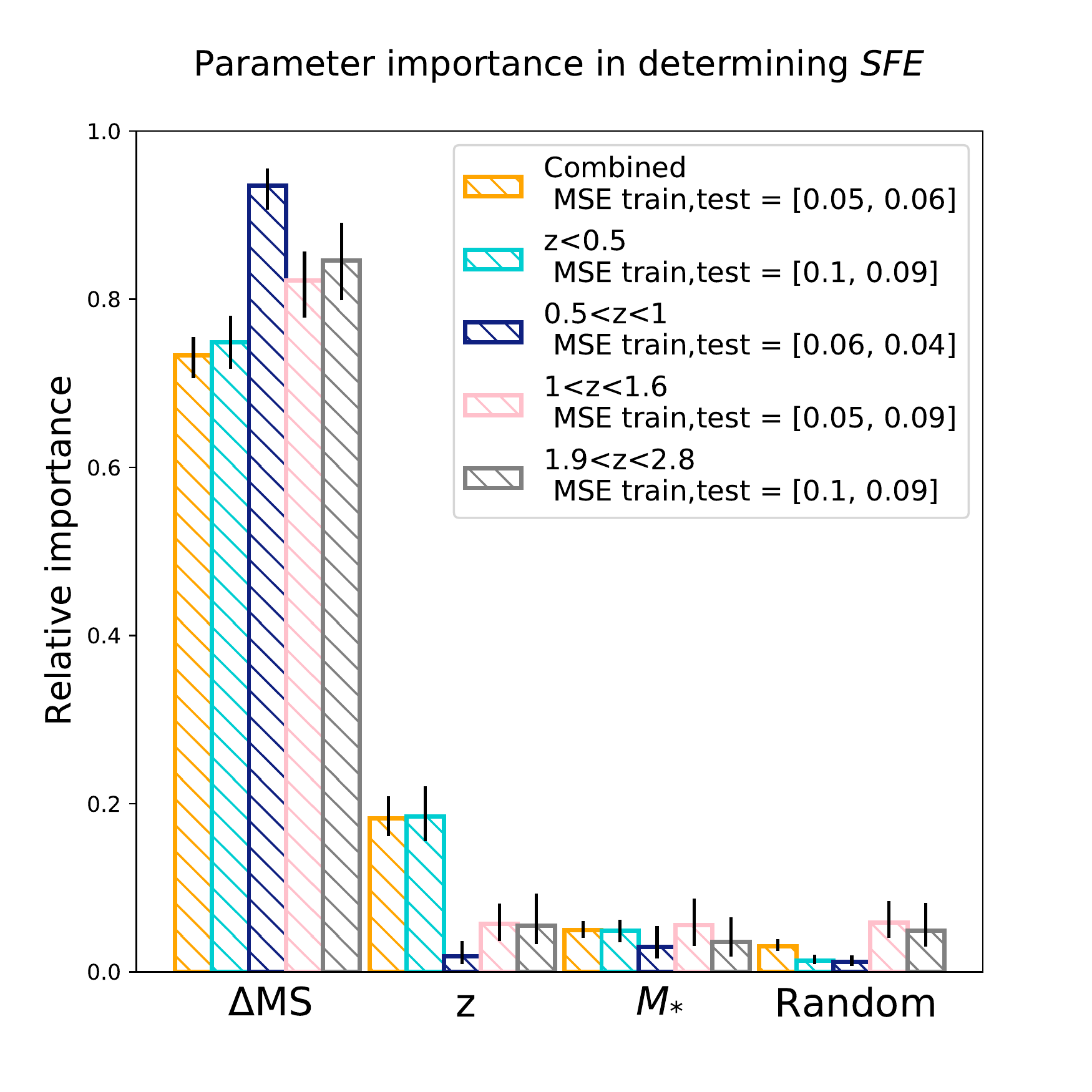}
    \caption{Random forest regression parameter importances in determining the star formation efficiency (SFE). The features considered are the distance from the star forming main sequence ($\Delta$ MS), the redshift (z), the stellar mass ($M_*$) and a control uniform random variable. In addition we split the sample into four redshift bins and include the results for the combined sample (orange). The figure shows that the SFE appears to primarily depend on the distance from the main sequence with a secondary dependence on redshift.}
    \label{fig:SFE_predict}
\end{figure}

In this subsection we compare our results to some 
well-known results in the literature. In particular \cite{Tacconi2020} and \cite{Genzel2015ApJ...800...20G} use a framework whereby they investigate the depletion time $\rm t_{dep}=1/SFE$ in galaxies. \cite{Tacconi2020} suggest that the depletion time depends primarily on redshift and that once the redshift is fixed it depends upon the distance of the galaxy from the Main Sequence. This is a different complementary approach with respect to the approach adopted by us. However, before discussing the approach, we first verify that our random forest methodology gives a result consistent with that found by
\cite{Tacconi2020} and \cite{Genzel2015ApJ...800...20G}.

Figure \ref{fig:SFE_predict} shows the importance, as inferred from the Random Forest, in determining the SFE (i.e. 1/$\tau$) for the following parameters: the distance from the main sequence ($\Delta$MS, where the main sequence is taken from \citealt{Renzini+PengMS2015ApJ...801L..29R}), the redshift (z), the stellar mass ($M_*$) and the (control) uniform random variable. 
We show the analysis for four different redshift bins and a combined bin (orange), i.e. all redshifts. The combined bin is present to ensure we are not excluding importance on redshift via the binning.
The Figure shows that the distance from the main sequence ($\Delta$MS) appears to be the most important parameter in determining the star formation efficiency, with redshift playing a secondary role (at least for the combined redshift bin), while stellar mass and the uniform random variable are both unimportant. Clearly, our methodology, when applied to this approach, provides results that are in full agreement  with the results of \cite{Tacconi2020} in finding that both $\Delta$MS and redshift appear to be the most important parameters in driving the SFE.

However, we {\rm note}
 that this process approaches the problem from the opposite way round, since SFE (and, therefore, depletion time) should be considered as a parameter contributing to the SFR rather than the other way round (where the information on star formation rate is included in the distance from the main sequence). Essentially, this approach inverts cause and effect, hence making it somewhat harder to uncover the intrinsic dependencies.

The additional problem of using $\Delta MS$ as a parameter to explore galactic properties, is that it automatically assumes that the SFMS is a fundamental, reference relation. However, we have demonstrated that the SFMS is not a fundamental relation, but a secondary byproduct of more fundamental relations (the SK and the MGMS). Hence, other parameters that are derived from it, such as $\Delta MS$, are even less fundamental, i.e. third-order byproducts of the more fundamental relations. Therefore, such tertiary parameters inferred from the SFMS, such as $\Delta MS$, should be used with caution.

 Summarising, our results are fully consistent with \cite{Tacconi2020}. Yet, we caution about exploring the SFE as a function of $\Delta MS$ if the goal is to investigate the fundamental relations and processes driving star formation in galaxies. However, the approach and scaling relations obtained by \cite{Tacconi2020} are extremely valuable for inferring quantities such as depletion time (i.e. 1/SFE) or molecular gas content, when these quantities cannot be measured directly.

\subsection{The role of AGN feedback in the local universe}

The properties of molecular gas in AGN has been explored previously in many works on both global and resolved scales \citep[][]{Maiolino1997ApJ...485..552M,2017XCOLDGASSAintongeApJS..233...22S, Ellison_AGN_2021MNRAS.505L..46E, 2021KossBATApJS..252...29K, Salvestrini_agn_2022A&A...663A..28S}. \cite{Ellison_AGN_2021MNRAS.505L..46E} used IFU data from EDGE-CALIFA and found evidence that AGN can deplete molecular gas in the central regions of their host galaxies. \cite{Maiolino1997ApJ...485..552M}, \cite{2021KossBATApJS..252...29K} and \cite{Salvestrini_agn_2022A&A...663A..28S} found that, on global scales, there is no evidence for depletion of molecular gas mass in AGN compared to star forming galaxies. These works support the idea that AGN can affect the molecular gas content of central regions of the AGN without necessarily affecting the total amount of molecular gas mass of the host galaxy, i.e. AGN feedback would work on local resolved scales not global. Our finding that AGN host galaxies follow an (integrated) AGN host galaxy MGMS, is generally consistent with these findings.

Our additional result that the (instantaneous) AGN luminosity does {\it not} affect the star formation rate, once the molecular gas content is taken into account, goes further in the direction that the instantaneous AGN feedback does not play a significant role on the properties of their host galaxies.
This result is interesting as there are multiple potential effects that might be associated with instantaneous AGN activity (traced by its luminosity) that are thought to impact the gas content in a galaxy or the rate with which it forms stars (such as ejection, shocks and photoionisation heating, turbulence), yet it would appear that these do not have a significant effect on determining the star formation rate or the content of molecular gas.

Our results agree with the work of \cite{Piotrowska2022MNRAS.512.1052P} where they analyse cosmological simulations to find that integrated power output of the AGN (i.e. integrated AGN feedback as traced by black hole mass) is much more important in quenching, i.e. impacting star-formation, than instantaneous power output (where instantaneous feedback is traced by AGN luminosity). Their interpretation is that the heating of the halo resulting from the integrated AGN activity prevents gas accretion on the galaxy and, therefore, delayed galaxy quenching as a consequence of starvation. To further test this scenario, an interesting avenue to explore with larger data sets would therefore be what the effect is of including black hole masses into the partial correlation coefficient and random forest regression analysis.

It is also important to note that these results are only valid in the local universe, and are not necessarily the case at high redshift \citep[][]{Circosta_2021A&A...646A..96C}. This will be an important area for further research once larger surveys are available.

\section{The SFMS is not a fundamental relation, but still very useful}

Throughout this paper, and also in \cite{Baker2022MNRAS.510.3622B}, we have extensively demonstrated that the SFMS, or its resolved version, the rSFMS, is not a fundamental relation. Is it therefore useless? Should we ignore it? In principle, we should explore galaxy properties by using the two more fundamental relations, i.e. the SK and the MGMS. However, both these relations imply the measurement of the molecular gas, either integrated or resolved, which is still challenging to obtain for large samples. Despite the major advances in the development of new millimetre facilities (e.g. ALMA), molecular gas masses are only available for hundreds of galaxies (possibly more than a thousand by combining different samples), and the statistics are even poorer (and samples more biased) at high redshift. On the contrary, the SFR can be measured with multiple tracers for millions of galaxies. Therefore, although the SFMS is an indirect, secondary relation, it is a very useful proxy of the two more fundamental relations, the SK and the MGMS, for large samples of galaxies, especially at high redshift.

\section{Conclusions}

By using a large sample of galaxies for which integrated gas content measurements are available we have explored the scaling relations between Star Formation Rate (SFR), Stellar Mass ($M_*$), and Molecular Gas Mass ($M_{H2}$), locally, at high redshift and for AGN host galaxies. We employ rigorous statistical tools to distinguish between direct, primary correlations from correlations that are an only indirect consequence of other more fundamental dependencies. In particular, we use Partial Correlation Coefficients and the Random Forest Regression Analysis to identify those relationships that are fundamental and those that are a by-product.

Our results can be summarised as follows

\begin{itemize}
    
    \item We find the existence of a tight correlation between molecular gas mass and stellar mass, the integrated version of the Molecular Gas Main Sequence, both locally and at high redshift.
    \item We confirm that the primary driver of SFR for galaxies both locally and up to redshift z=2.8 is the molecular gas mass of the galaxy.
    \item Once the dependence on molecular gas is taken into account, there is no residual dependence of the SFR on the stellar mass, both locally and at high-z.
   This finding implies that the SFMS is not a fundamental scaling relation, rather it is an indirect by-product of the SK and MGMS relations. This is in agreement with what was found locally on resolved scales. 
    \item We provide a parametrisation of the MGMS both in the local universe and for three different bins of redshift. We find that $M_{H_2}$ is offset to higher values for a fixed $M_*$ as redshift increases. We offer a parameterisation of this evolution of the MGMS, which peaks at z$\sim$2.3, i.e. around cosmic noon.
    \item We also explore and parametrise the redshift evolution of the Star Formation Efficiency, $\rm SFE=SFR/M_{H2}$.
    \item We show that the observed evolution of the SFMS with redshift is simply a consequence of the redshift evolution of the MGMS and SFE, although the primary driver of the SFMS redshift evolution is the MGMS.
    \item The molecular gas mass is the primary driver of the SFR of galaxies in the local universe with SFE playing a secondary, but important role. SFE has greater importance at z$>$0.5, but it is still primarily molecular gas mass that determines the SFR.
    \item Galaxy morphology (B/T) does not directly contribute to determining the SFR for star-forming galaxies. 
    \item We find that the same findings hold also for AGN host galaxies, i.e. also in AGN host galaxies the molecular gas and stellar mass are correlated through a MGMS,  that the SFR is primarily driven by the gas content with no direct correlation with stellar mass, hence implying that the SFMS is not a fundamental relation, but a by-product of the SK and MGMS.
    \item Furthermore, we find AGN luminosity does not play a role in determining the SFR of the AGN host galaxy. This indicates that direct, instantaneous AGN feedback does not play a significant role in quenching star formation in their host galaxies (at least on global, galactic scales), although it can have an important role in terms of delayed, preventive feedback.

\end{itemize}

\section*{Acknowledgements}

W.B., R.M., and A.B. acknowledge support from the ERC Advanced Grant 695671, “QUENCH” and from the Science and Technology Facilities Council (STFC).
R.M. also acknowledges funding from a research professorship from the Royal Society.
L.L. thanks support by the Academia Sinica under the Career Development Award CDA107-M03 and the Ministry of Science \& Technology of Taiwan under the grant MOST 111-2112-M-001-044 -.
HAP acknowledges support by the National Science and Technology Council of Taiwan under grant 110-2112-M-032-020-MY3

\section*{Data Availability}

The ALMA data used is publicly available through the ALMA archive http://almascience.nrao.edu/aq/.
The MPA-JHU catalogue is publicly available at https://wwwmpa.mpa-garching.mpg.de/SDSS/DR7/.
The remaining datasets used for this analysis are all publicly available and attached to their survey papers. These are: MASCOT \citet{DominikaMascott2022MNRAS.510.3119W}, ALLSMOG \cite{2017CiconeALLSMOGA&A...604A..53C}, xCOLD GAS \cite{2017XCOLDGASSAintongeApJS..233...22S}, EDGE \cite{2017BolattoEDGEApJ...846..159B}, HERSCHEL \cite{2018BertemesMNRAS.478.1442B}, ALMaQUEST \cite{Lin2020}, BAT \cite{2021KossBATApJS..252...29K}, and PHIBBS \cite{Tacconi_PHIBBS_2018ApJ...853..179T}.



\bibliographystyle{mnras}
\bibliography{example} 




\appendix

\section{Redshifts}

\begin{figure}
    \centering
    \includegraphics[width=\columnwidth]{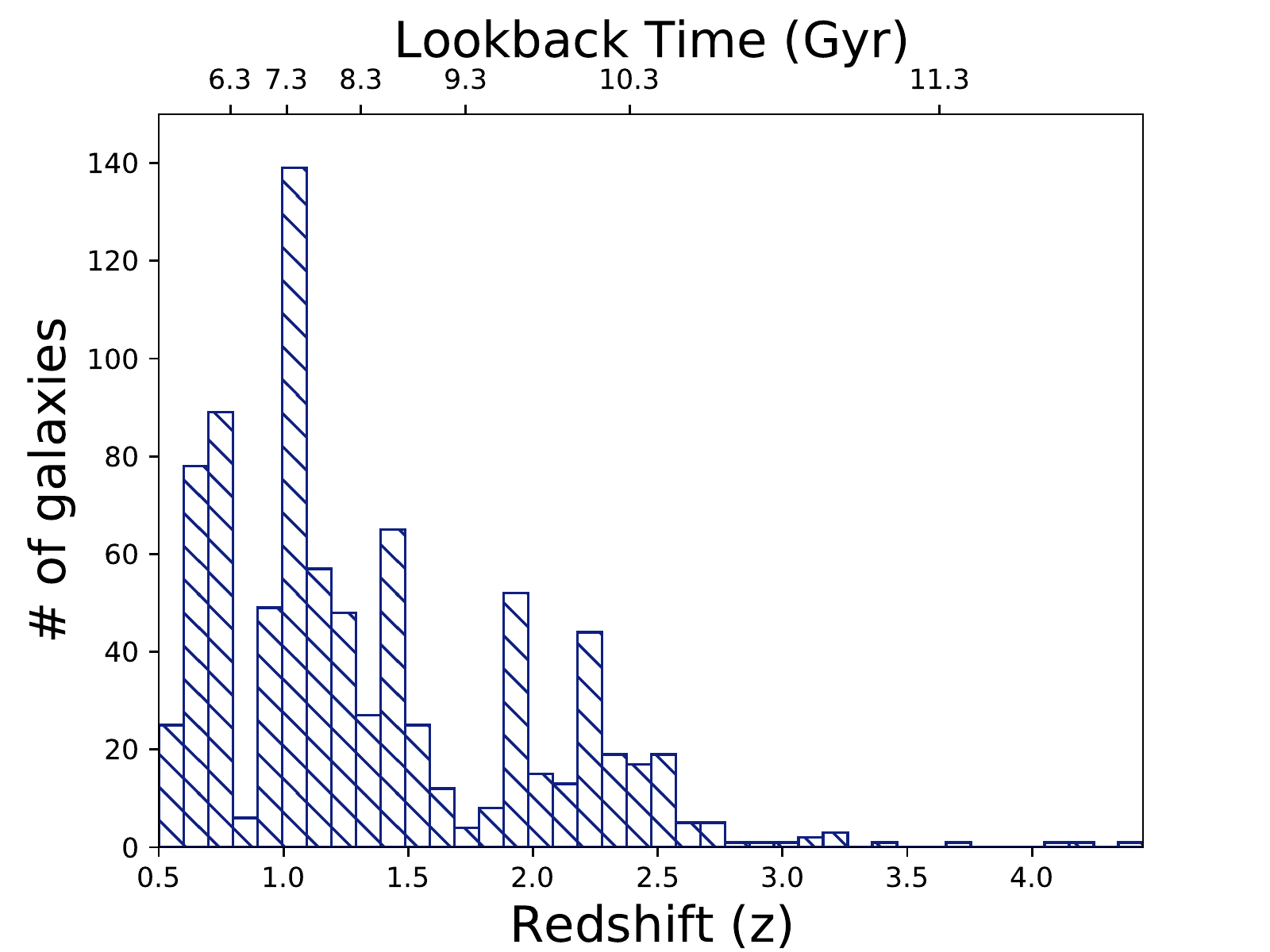}
    \caption{Histogram of galaxy redshifts in the PHIBBS sample.  The distribution is tilted towards lower redshifts. This explains why we primarily select lower redshift bins in the range z=0.5-2.8.}
    \label{fig:redshifts}
\end{figure}

We present in this Appendix Figure \ref{fig:redshifts}, a histogram containing the redshifts from the used sub-sample of the PHIBBS galaxies. From this histogram it becomes apparent why we selected the three bins of redshift we did. The aim was to have approximately equal numbers of galaxies in each bin and for each bin to be a similar size. This of course cuts down the maximum redshift of the largest z bin due to the small number of very high redshift galaxies.

\section{Effects of metallicity-dependent conversion factor in local universe}

In section \ref{sec:localSF} we explored which is the dominant driver out of molecular gas mass and stellar mass in determining the star formation rate of star forming galaxies in the local universe. In doing so we used a constant conversion factor ($\rm \alpha_{CO}$) as given by \cite{Bolatto2013}. It is strongly suspected that $\rm \alpha_{CO}$ has an intrinsic dependence on metallicity \citep[][]{Accurso2017MNRAS.470.4750A}. Here we include the results of applying a metallicity-dependent conversion factor from CO to H$_2$. The conversion factor we use is given in section \ref{sec:methods}.  

An initial expectation would be that we would not expect it to cause a significant difference, which would mirror what was found by \cite{Baker2022MNRAS.510.3622B}, where we used the same methods (partial correlation coefficients and random forest regression) on resolved scales and found that both the constant and metallicity dependent conversion factor scenarios gave almost the same results. 

In order to obtain consistent metallicity measurements for the different galaxies that make up our local star-forming sample, we estimate them using the fundamental metallicity relation parameterisation given in \citet{2020CurtiMNRAS.491..944C}. This enables us to predict the metallicity of each galaxy from its stellar mass and star formation rate. We note that, as we are using stellar masses and star formation rates from the MPA-JHU catalogue (making our sample consist purely of galaxies observed in SDSS DR7), this parameterisation should give accurate metallicities for our galaxies as \citet{2020CurtiMNRAS.491..944C} found a residual scatter of just 0.054dex for the global galaxy population around their FMR parameterisation.

\begin{figure*}
    \centering
    \includegraphics[width=1.1\columnwidth]{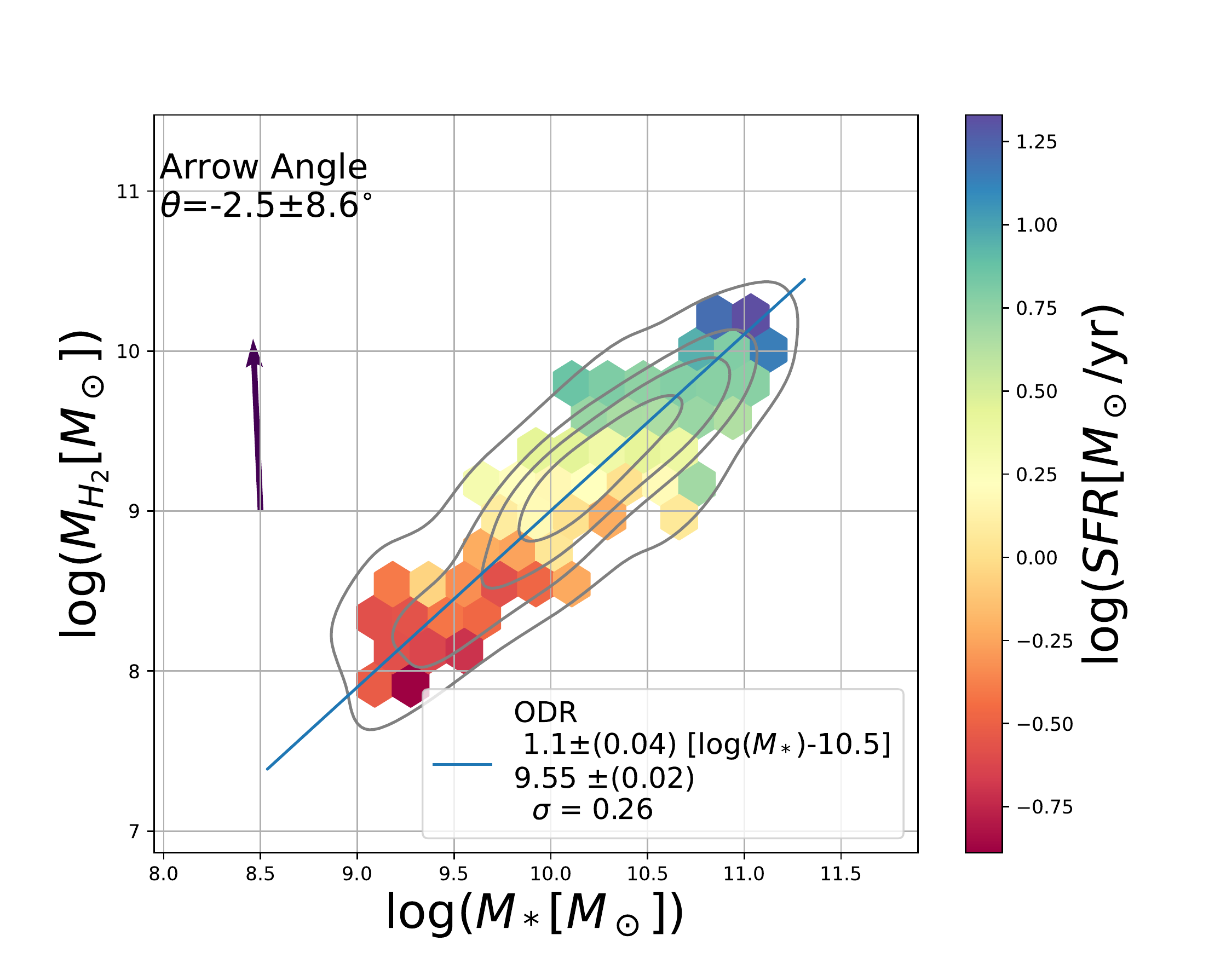}
    \includegraphics[width=0.9\columnwidth]{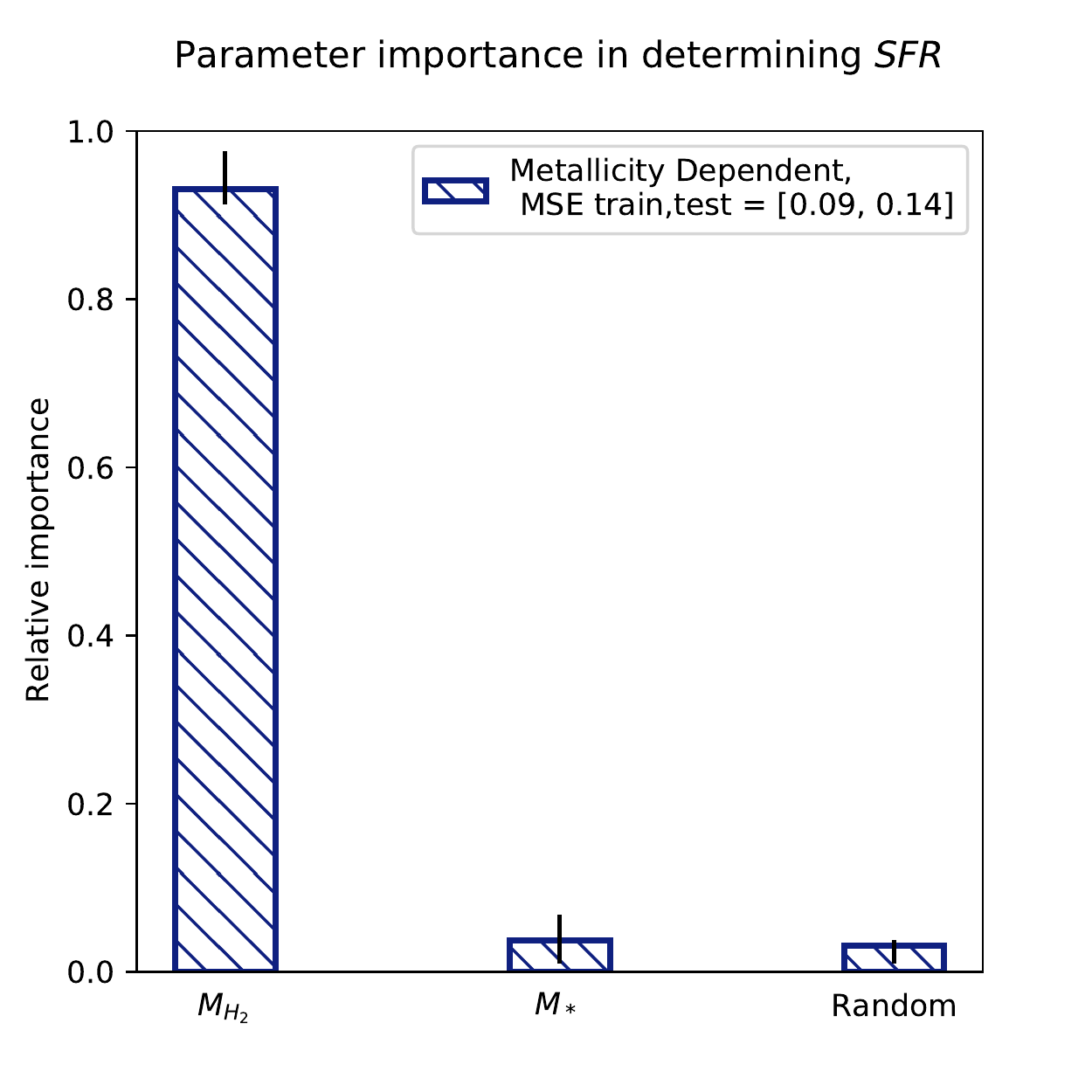}
    \caption{The same plots as Figures \ref{fig:Mstar-H2-SFR-Starforming} \& \ref{fig:RF_starforming}, only this time with M$_{H2}$ calculated from CO with a metallicity dependent $\rm \alpha_{CO}$. Left: 2D histogram of molecular gas mass against stellar mass colour coded by the mean log SFR in each bin. Density contours describe the distribution of galaxies with the outer contour enclosing 90\% of galaxies. The arrow points in direction of steepest gradient in SFR. The ODR line shows a tight MGMS. 
    Right: random forest regression parameter importances in determining the star formation rate. The parameters evaluated are molecular gas mass, stellar mass and a uniform random variable. These plots show how using the metallicity-dependent conversion factor does not significantly change any of our results or conclusions in section \ref{sec:localSF}. The star formation rate is still found to be totally driven by the molecular gas mass with little dependence on stellar mass. This means that the SFMS is not a fundamental scaling relation. }
    \label{fig:zdep}
\end{figure*}

Figure \ref{fig:zdep} contains the same plots as in Figures \ref{fig:Mstar-H2-SFR-Starforming} \& \ref{fig:RF_starforming}, only this time for the metallicity-dependent conversion factor scenario. This enables us to check whether our previous results are affected by the conversion factor used. We immediately see that the results are consistent with those found for the constant conversion factor case presented in section \ref{sec:localSF}. The star formation rate is almost entirely determined by the molecular gas mass with little dependence on the stellar mass. 
In addition we find the form of the MGMS, defined by the ODR line in the left plot of Figure \ref{fig:zdep}, for the metallicity-dependent $\rm \alpha_{CO}$ is very similar to that found for the constant $\rm \alpha_{CO}$ case.
Therefore we find, as was found by \citet{Baker2022MNRAS.510.3622B} on resolved scales, that for analysis of the SFR dependence, the difference between using a constant $\rm \alpha_{CO}$ and one that depends on metallicity does not significantly affect our results.


\bsp	
\label{lastpage}
\end{document}